\documentclass{aa}

\usepackage{graphicx}
\usepackage{txfonts}
\usepackage{amsmath}
\usepackage[]{hyperref}
\usepackage{ulem}
\providecommand{\kms}{\mathrm{km\,s^{-1}}}
\begin{document} 

   \title{Correcting the fiber-aperture bias affecting galaxy stellar populations in the Legacy\ Sloan Digital Sky Survey}
   \subtitle{Aperture corrections to absorption indices based on CALIFA integral field observations}

\titlerunning{Spectroscopic aperture effects in extragalactic stellar populations in the SDSS}
   \author{Stefano Zibetti,
          \inst{1,2}
          Jacopo Pratesi,
          \inst{1,2}
          Anna R. Gallazzi,
          \inst{1}
          Daniele Mattolini
          \inst{3,1}
          \and
          Laura Scholz-D\'iaz
          \inst{1}
          }
\authorrunning{Zibetti et al.}
   \institute{INAF-Osservatorio Astrofisico di Arcetri, Largo Enrico Fermi 5, I-50125 Firenze, Italy\\
              \email{stefano.zibetti@inaf.it}
        \and
             Dipartimento di Fisica e Astronomia, Universit\`a degli Sudi di Firenze, Via G. Sansone 1, I-50019 Sesto Fiorentino, Italy
        \and
            Dipartimento di Fisica,
            Universit\`a di Trento, Via Sommarive 14, I-38123 Povo (TN), Italy
             }

   \date{Received 28 August 2025 / Accepted 15 December 2025}

  \abstract
   {A detailed characterization of stellar population properties is crucial for understanding galaxy evolution. Their inference for statistically representative samples requires deep multi-object spectroscopy, typically obtained with fiber-fed spectrographs integrating only a fraction of galaxy light. The Legacy Sloan Digital Sky Survey (SDSS I-II) represents the most studied local Universe dataset in this context. Its\ fibers typically collected $\sim 30$\% of total flux. Ubiquitous stellar population gradients systematically bias SDSS measurements toward central properties by an as-yet-unquantified amount.}
  {Our aim is to quantify spectroscopic fiber aperture effects for a representative sample of local Universe galaxies over the SDSS redshift range, and to devise empirical recipes to correct stellar absorption indices from SDSS fiber spectra to aperture-free total values.}
   {We leveraged CALIFA integral-field spectroscopy to simulate fiber-fed observations at redshifts $z = 0.005$--$0.4$, accounting for seeing effects. We analyzed systematic aperture correction trends across galaxy morphologies and derived correction recipes based on fiber-measured indices, the global $g\!-\!r$ color, the absolute $r$-band magnitude $M_r$, and the physical half-light radius $R_\mathrm{50}$.}
   {Corrections for absorption indices typically reach $\gtrsim 15$\% of their dynamical range at $z \sim 0.02$, decreasing to $\sim 7$\% at $z \sim 0.1$ (median SDSS redshift) and becoming negligible above $z \sim 0.2$. Spiral galaxies exhibit the largest aperture effects due to their strong internal gradients. Our correction recipes, split into first-order terms (from fiber-indices and $g\!-\!r$ color) and second-order terms (from $M_r$ and $R_\mathrm{50}$) and applied to the SDSS-DR7 dataset, significantly reduce the scatter in stellar population diagnostic planes
   and enhance the bimodality in age-sensitive diagrams. The corrections reveal systematic overestimates of old galaxy fractions by up to 10\% and an underestimate by $\gtrsim 0.2$\,mag of the transition luminosity at which old galaxies become dominant.}
   {Aperture corrections significantly impact the observational tracers of stellar populations from fiber spectroscopy and tighten correlations between stellar population properties. Absorption index corrections applied to the SDSS-DR7 dataset will provide a robust local benchmark for galaxy evolution studies.}


   \maketitle
%
\section{Introduction}\label{sec:intro}

Along the evolutionary history of a galaxy, different generations of stars, either formed in situ or accreted in merger events, keep  accumulating and preserving the archaeological records of the physical conditions at their birth. The optical spectrum of a galaxy encodes all this information. While rough estimates of the age of the dominant component of a galaxy are possible by means of basic diagnostics such as broadband colors or the strength of prominent absorption features (e.g., the 4000\,\AA~and high-order Balmer lines), accurate estimates of the ages and chemical composition of different generations of stars are accessible thanks to a wealth of relatively high-resolution high signal-to-noise ratio (S/N) spectra to be confronted with reliable spectral models. In fact, since the introduction of the stellar population synthesis (SPS) technique by \cite{tinsley_1980}, vast progress has been made in all the ingredients required to properly link a stellar population's physical properties with the observational spectral characteristics, such as stellar evolutionary tracks, isochrones, stellar atmospheres, and spectral templates \citep[see][for a review]{conroy_2013}.

Despite the uncertainties in the SPS models that propagate into the estimated properties \citep[see, e.g.,][]{conroy+09} and, combined with different inference methods, that can yield systematic differences of approximately 0.1 dex between different works, the general trends of the so-called stellar-population scaling relations are well established in the $z\sim 0$ Universe. The most massive galaxies exhibit the highest mean stellar ages and metallicities, and vice versa, on average, the least massive ones are young and metal-poor \citep[e.g][]{gallazzi+05,Gallazzi:2021aa,2008MNRAS.391.1117P,Trussler:2020aa,Neumann.2021}. The exact shape of these relations and the partition of galaxies in different age and star-formation activity regimes (often referred to as bimodality) appears to be much more model- and method-dependent \citep[see, e.g.,][]{Mattolini:2025}. However, these relations are fundamental benchmarks for galaxy evolution models, and they provide the essential zero-point for tracing the empirical evolution of stellar population propertied across cosmic time \citep[e.g.,][]{Sommariva+2012,2014ApJ...788...72G,Cullen+2019,Kriek+2019,Beverage+2023,Saracco+2023,Bevacqua+2024,Gallazzi_paperI:2025,Gallazzi_paperII:2025}.

A thorough statistical characterization of the stellar population scaling relations relies on the availability of a large database of spectroscopic observations over a broad wavelength range in the rest-frame optical, with moderate spectral resolution, and relatively high S/N (i.e. $\mathrm{S/N}\gtrsim 10-20$). This is achievable only through highly multiplexed multi-object fiber-fed spectrographs. In the low-$z$ Universe, the Legacy Sloan Digital Sky Survey \citep[SDSS I and II, see][]{SDSS, EDR, strauss_etal02} was optimally designed to fully satisfy these requirements. Thanks to its data-quality homogeneity and its extensive dataset, it still represents the current benchmark in this context.\ However, despite the generous 3 arcsecond diameter aperture of the SDSS spectroscopic fibers, typical spectra collect only $\sim 30\%$ of the galaxy light on average \citep{gallazzi+05}, preferentially sampling the central regions. Given the well-established differences in stellar populations between disks and classical bulges or ellipticals \citep[e.g.,][]{peletier_balcells_1996}, the systematic morphological differences between galaxies of different masses \citep[e.g.,][]{nair_abraham_2010}, and the existence of stellar population gradients \citep[e.g.,][]{Gonzalez-Delgado:2014aa}, this aperture bias is expected to affect the general characterization of stellar populations and their scaling relations. Although the amplitude of this bias has long been debated, currently no quantitative assessment is available in the literature for stellar population properties.\footnote{Aperture corrections for emission lines and, more specifically, for star-formation rate estimates have been computed and applied in the past decades, for example by \cite{brinchmann+04} and updates following \cite{Salim:2007aa}, and by \cite{2013A&A...553L...7I}.}

The direct way to get around the aperture bias is to employ integral field spectroscopy (IFS). This technique allows us to map and integrate the light over a much larger extent than a single fiber, but this comes at the price of a multiplication of spectra per object, which enforces a trade-off between sample size, spatial (or spectral) resolution, and spatial coverage. In fact, the major IFS galaxy surveys available at the moment offer a much more limited sample size with respect to the Legacy\ SDSS single-fiber survey, and, in addition, only for a portion of the sample the spatial coverage extends to two effective radii or beyond, so as to make aperture effects negligible. In the case of SAMI \citep{Bryant:2015,Croom+2021} the number of galaxies observed over such a large footprint is $\lesssim 1\,000$, while in the case of SDSS-MaNGA \citep[][]{Bundy:2015} this number is $\sim 3\,900$. Quite obviously, when compared to the number of galaxies available in the Legacy\ SDSS ($\gtrsim 354\,000$ galaxies with $\mathrm{S/N}\geq 10$, see below), these differences in sample size preclude a number of multivariate analyses.

In order to preserve the sample size of the Legacy\ SDSS, the alternative solution is to attempt to correct the SDSS spectroscopic measurements for the aperture biases on a statistical basis. This is the approach that we took for this work. We employed IFS data to simulate what the SDSS spectrum would look like for different redshifts of a given galaxy. Its spectral properties were then compared with the properties of the spectrum integrated over the full galaxy extent. By correlating the differences with observational properties that can be consistently and easily obtained over a broad redshift interval, our aim was to establish correction recipes that can be applied to the full SDSS galaxy sample. As we show in Sect. \ref{sub:CALIFA}, the Calar Alto Legacy Integral Field Area (CALIFA) survey \citep{Sanchez:2012aa} turns out to be the ideal IFS survey for these goals, mainly due to its large spatial coverage and relatively high physical resolution, despite its relatively small sample size.
Our analysis focuses on spectral indices as tracers of stellar population physical properties. Their ability to condense very efficiently such information in a small set of numbers that are well defined and independent of the flux normalization makes them ideal quantities to compute aperture corrections, as opposed to full spectra.

In Sect. \ref{sec:spec_simul} we describe how we simulate SDSS-like fiber spectra using CALIFA IFS for a representative sample of galaxies at different redshifts. We also illustrate and quantify the fiber losses and the typical biases affecting different classes of galaxies, as a function of redshift, for the most commonly used stellar indices. In Sect. \ref{sec:apercorr_recipes} we introduce the procedure to estimate the aperture effects on absorption indices as a function of observable parameters (redshift, measured index, global color, luminosity, and size), based on the analysis of the CALIFA sample. The application of these recipes to correct the fiber-aperture indices of the SDSS-DR7 galaxy sample (825\, 263 unique galaxies in total, more than 354\, 000 with $\mathrm{S/N}\geq 10$) to integral-light is presented in Sect. \ref{sec:SDSScorrected}. We analyze how the distributions of galaxies in index-index diagnostic planes are modified by these corrections. In particular, we show how important these corrections are in quantifying the bimodality between active star-forming galaxies on one side, and old or passive galaxies on the other, as a function of luminosity. A summary and concluding remarks are provided in Sect. \ref{sec:conclusions}. Throughout the paper a standard flat Lambda cold dark matter ($\Lambda$CDM) cosmology with $h=0.7$ and $\Omega_\Lambda = 0.7$ is adopted. For conciseness, we use SDSS to refer exclusively to the Legacy SDSS (I and II), whose final data release is DR7 \cite{SDSS_DR7}.

\section{Modeling the aperture biases of fiber spectra from IFS data cubes}\label{sec:spec_simul}
In this section we describe how IFS data cubes from the CALIFA survey are used to generate a
comprehensive library of simulated fiber-spectra for a broad range of galaxies' properties
(both extensive, such as mass and effective radius, and intensive, such as morphology and specific SFR)
and a range of redshift, which are representative of the SDSS galaxy dataset in the local Universe,
specifically of the SDSS Main Galaxy sample \citep{strauss_etal02}.
Essentially, the idea is to start from an IFS observation at high spatial resolution, degrade it to the resolution at which SDSS would see that galaxy at a given redshift $z$ and integrate the light that falls within the 3-arcsec-diameter aperture of an SDSS fiber. For each galaxy, from the comparison of the observational properties (chiefly absorption indices) of these simulated spectra with those of the spectrum integrated over the full area of the galaxy (full-aperture spectrum) we  then derive the estimates of the aperture biases relevant for the characterization of the stellar population properties.
In Sect. \ref{sub:CALIFA} we describe the CALIFA dataset; the method to simulate fiber spectra through SDSS-like circular apertures is detailed in Sect. \ref{sub:apersims}; in Sect. \ref{sub:CALIFA_fluxfrac} we quantify the fraction of light collected by the fibers as a function of redshift and general galaxy properties (morphology, luminosity) and in Sect. \ref{sub:CALIFA_aperbias} we illustrate the resulting biases in terms of spectral shape and absorption index strengths.

\subsection{The CALIFA dataset}\label{sub:CALIFA}
In order to model the aperture biases of SDSS galaxy spectra for the main stellar absorption indices of interest over the visible range, we need IFS observations of a representative sample of galaxies in the nearby Universe providing: (i) extended spectral coverage in the visible, with a spectral resolution of $\lesssim 100\,\mathrm{km~s}^{-1}$; (ii) extended spatial coverage to include $\gtrsim 2.5 \mathrm{R_e}$; (iii) good angular resolution to enable the simulation of a 3$^{\prime\prime}$ aperture on a galaxy at redshifts as low as $\sim 0.01$, which corresponds to a physical resolution of $\lesssim 1$ kpc. We note that most SDSS galaxies with spectroscopic $\mathrm{S/N}\geq 10$ have $\mathrm{R_e}$ between 1.5 kpc and 5 kpc (see, e.g., Fig. \ref{fig:corr_flow_HdHg}).

The Calar Alto Legacy Integral Field Area (CALIFA) survey \citep{Sanchez:2012aa} represents an excellent match to all three requirements. This makes it preferable despite the larger sample sizes of surveys such as MaNGA and SAMI, as we discuss in the following paragraphs. For this work we relied on the CALIFA sample of COMBO data cubes, which are part of the final public data release \citep[DR3]{Sanchez:2016aa}. These are all the galaxies belonging to the CALIFA Main Sample \citep{Walcher:2014aa} that have been observed in both the medium-resolution blue setup (V1200) and the low-resolution red setup (V500) and for which the combined and homogenized low-resolution data cube (COMBO) is available. The sample is effectively a random draw from the size-selected main sample, which was defined with the following properties: isophotal $r$-band diameter $45^{\prime\prime}\,<\,\mathrm{iso}A_r < 79.2^{\prime\prime}$ from the SDSS DR7 \citep{SDSS_DR7} photometric galaxy catalogue; available redshift in the nominal range 0.005 $< z <$ 0.03; good photometric quality and visibility from the Calar Alto Observatory. 
As noted in \cite{Walcher:2014aa}, the representativeness of the sample is good enough to compute volume corrections to a volume limited sample. However, it should be noted that dwarf galaxies ($M_{r,\mathrm{abs}}>-19~\mathrm{mag_{AB}}$, especially of late morphological types) are under-represented.
We verified that the CALIFA sample provides an excellent coverage of the luminosity distribution of the overall SDSS sample (limited to $\mathrm{S/N}\geq 10$, see Sect. \ref{subsec:SDSSsample} for its definition and characterization).

The COMBO cubes cover a useful spectral range between 3700\,\AA~and 7140\,\AA, which includes all main stellar absorption features in the optical, except the red-most ones (e.g., the Ca-Triplet). The spectral resolution of 6\,\AA~FWHM is good enough to simulate the majority of SDSS galaxies in the Main Galaxy sample, above the corresponding velocity dispersion of $\sim 120$ km s$^{-1}$ (depending on the exact reference wavelength). 

The 2D PSF of the data cubes on the sky plane has a typical FWHM of $\sim 2.6$~arcsec (corresponding to 0.25 to 1.5 kpc over the redshift range of the CALIFA sample), which essentially enables a reliable simulation of the light collected in an SDSS fiber as long as the target redshift of the simulation is larger than the native redshift of the galaxy. We find that  $>97.3$\% of the SDSS galaxies considered for stellar population analysis \citep[i.e., selected by $\mathrm{S/N}>10$ and $z>0.005$, e.g., ][]{Mattolini:2025,gallazzi+05} are above the maximum redshift of the CALIFA sample. We can thus cover most of the relevant parameter space with reliable simulations, with the only exception of the very low-redshift tail of the SDSS distribution.

 It should be noted that in the case of MaNGA one should restrict to the secondary sample ($3\,724$ galaxies) in order to cover $\gtrsim 2.5 \mathrm{R_e}$, sacrificing spatial resolution to worse than $\sim 2$\ kpc for the vast majority of the sample,\footnote{See figure 1 on \url{https://www.sdss4.org/dr14/manga/manga-target-selection/} \citep{Wake:2017}.} which is inadequate for our purposes. Similarly, SAMI ($2\,100$ galaxies) privileged sample statistics over spatial resolution, with a much poorer sampling \citep[61 fibers per galaxy, as opposed to 331 in CALIFA;][]{Bryant:2015} and much lower effective resolution. In fact, from \cite{Croom+2021} we can estimate that the effective radius of galaxies above $\sim 10^{9.5}\,\mathrm{M_\odot}$ (i.e., the mass range relevant to SDSS stellar population analysis) projects to between 1 and 5 arcseconds, thus between $\sim 0.5$ and 3 times the PSF FWHM. This is at least a factor of 3 worse than CALIFA, for which the average ratio between the effective radius and the PSF FWHM ranges approximately between 5 and 10 (16th--84th percentiles). 

Of the 396 available COMBO data cubes, two galaxies are excluded because of quality issues: UGC\,11694 because of a very bright star near the center, which contaminates a significant portion of the galaxy’s optical extent; UGC\,01123 because of artifacts in the noise spectrum \citep{zibetti+17}. 
Furthermore, as we  explain in Sect. \ref{sec:apercorr_recipes}, a reliable measurement of the Petrosian half-light radius $R_{\mathrm{Petro},50}$ is required to evaluate the second-order aperture corrections. As the standard measurements of $R_{\mathrm{Petro},50}$ in the DR7 catalogs are known to be biased for galaxies with a projected size of several tens of arcsecond, we decided to rely on the NASA-Sloan Atlas\footnote{\url{http://nsatlas.org}} (hereafter NSA). The NSA not only provides more reliable photometric quantities (including $R_{\mathrm{Petro},50}$), based on the techniques presented in \cite{Blanton_etal:2011}, but also includes accurate k-corrections to rest-frame bands based on the application of \texttt{kcorrect} \citep{Blanton_kcorrect} to the extended SED, ranging from the GALEX far UV to the SDSS $z$-band. Seven CALIFA galaxies do not have a match in NSA, therefore they are excluded. We further exclude four galaxies with $R_{\mathrm{Petro},50}$ corresponding to less than 1 kpc, because these measurements are both unreliable (e.g., by comparison with independent CALIFA estimates) and irrelevant for the SDSS dataset, which lacks such compact galaxies.

In order to measure the reference full aperture spectrophotometric quantities in a homogeneous way for all galaxies, we defined the footprint as the area within the isophote of $23.5~\mathrm{mag~arcsec}^{-2}$ in $r$-band.\footnote{The $23.5~\mathrm{mag~arcsec}^{-2}$ isophotes are computed on the corresponding SDSS $r$-band images, which provide a better sky subtraction, less noise and better photometric calibration with respect to the CALIFA data cubes at this surface brightness level. The isophotal contours are then translated to the data cubes. We note that by selection, this isophotal level is fully included in the CALIFA cubes.} This definition is justified by considerations on the finite spatial coverage and the noise properties of the CALIFA data cubes. In fact, on the one hand, the CALIFA coverage extends at most out to 2.5-3 half-light radii. On the other hand, integrating the signal of the cubes where the galaxy surface brightness is low, may introduce biases due to possible inaccurate sky subtraction, as well as lowers the overall S/N. The adopted footprint definition is a compromise between extending the coverage as far as possible and limiting the background noise contribution. Bright fore- and back-ground sources are masked out of the footprint as well.

We compare the flux in this footprint, $F_\mathrm{footprint}$, with the total magnitude obtained by the CALIFA collaboration on the SDSS $r$-band images using the growth curve (GC) extrapolation technique \citep{Walcher:2014aa}, and compute the footprint fraction as the ratio of the flux integrated in the footprint to the flux from the GC analysis:
\begin{equation}\label{eq:f_fp}
f_\mathrm{footprint}=\frac{F_\mathrm{footprint}}{F_\mathrm{GC}}.
\end{equation}

We further excluded eight galaxies with $f_\mathrm{footprint}<0.5$ or $f_\mathrm{footprint}>1.5$. The resulting distribution of $f_\mathrm{footprint}$ for the final sample of 375 galaxies has a median value of 0.89, and 16th and 84th percentiles of 0.74 and 1.02, respectively.\footnote{The $f_\mathrm{footprint}>1$ are accounted for by uncertainties on the sky background or by different masking of bright close-by sources, which are independently estimated in the two procedures of flux integration.} Due to the adopted cut in surface brightness, galaxies with a high Sersic index and galaxies with an overall low surface brightness have typically lower $f_\mathrm{footprint}$. In fact, looking at the distribution in $f_\mathrm{footprint}$ as a function of stellar mass, color and Hubble type, we find a median $f_\mathrm{footprint}\approx 0.75$ for elliptical galaxies and for $M_* \gtrsim 10^{11.2} \mathrm{M}_\odot$. At the opposite extreme of the Hubble sequence, roughly corresponding to galaxies with $M_*  \lesssim 10^{9.5} \mathrm{M}_\odot$, the median $f_\mathrm{footprint}$ is $\approx 0.8$. As a consequence, for these galaxies aperture corrections will be somewhat underestimated. Nonetheless, we note that these footprint fractions are significantly larger than the typical fractions of light in fiber of SDSS spectra, which is $\approx 0.3$ \citep[][and Sect. \ref{sub:CALIFA_fluxfrac}, Fig. \ref{fig:fiberfrac_distr}]{gallazzi+05}. 

\subsection{Simulating spectra through a circular aperture}\label{sub:apersims}
\subsubsection{Method}\label{subsub:apersim_method}
In order to simulate the spectrum that the SDSS would collect for a CALIFA galaxy once it is located at a given redshift $z$, we have to take into account (i) the area of the galaxy projected on the sky that falls inside the circular aperture of the fiber, and (ii) the blurring of the 2D distribution of the light as a consequence of the seeing. The relative scaling of the galaxy proper physical size to the angular size of the fiber aperture and the seeing radius (or, alternatively, the full width at half maximum FWHM of the point spread function PSF) depends on the angular diameter distance, which is a function of the redshift $z$ and of the adopted cosmology. To start with, the angular size $\theta_0$, in the observed frame of a CALIFA galaxy at $z=z_0$, that corresponds to the size $\theta_s$ in the frame of the simulated SDSS observation at $z=z_s$, is given by 
\begin{equation}
    \theta(z_0)=\theta(z_s)\frac{D_A(z_s)}{D_A(z_0)}\equiv \theta(z_s)\, {\mathcal{R}}_A(z_s,z_0),
\end{equation}
where $D_A(z)$ is the angular diameter distance of the chosen cosmology. We note that the angular distance ratio ${\mathcal{R}}_A(z_s,z_0)$ is independent of $H_0$ (or $h$). In the relatively low-redshift regime covered by our simulations, the other cosmological parameters ($\Omega_m$, $\Omega_\Lambda$) have only a mild effect of a few percents, as long as they are in the commonly accepted ranges.

From a purely logical point of view, the most straightforward way to proceed is the following:\\
1) For each wavelength slice in the CALIFA data cube, the image would be convolved with a kernel in order to reproduce the PSF in the simulated frame, i.e., of the observing conditions of SDSS. By assuming a simple Gaussian as a PSF for both CALIFA and SDSS, the convolution kernel is given by a Gaussian of width 
\begin{equation}\label{eq:sigma_conv}
    \sigma_{\mathrm{conv}}=\sqrt{\sigma_{\mathrm{SDSS}}^2\,{\mathcal{R}}^2_A(z_s,z_0)-\sigma_{\mathrm{CALIFA}}^2 }
\end{equation} (in arcseconds), in the native CALIFA observed frame, where $\sigma_{\mathrm{SDSS}}$ and $\sigma_{\mathrm{CALIFA}}$ are the $\sigma$ of the PSF Gaussians corresponding to FWHM of 1.6$^{\prime\prime}$ and 2.57$^{\prime\prime}$ for SDSS and CALIFA, respectively (see discussion in Sect. \ref{subsub:caveats}).\footnote{For a few galaxies the low-redshift tail of the simulations yield a negative argument for the square root. In these cases we drop the galaxy at that specific redshift, which is not considered in the corrections database.}\\
2) A circular mask of diameter $D(z_s)$ of $3^{\prime\prime}$ times the angular diameter distance ratio ${\mathcal{R}}_A(z_s,z_0)$ would be used to integrate the flux in each wavelength slice.

In practice, this approach proves to be very slow from a computational point of view, as it requires  convolving
each wavelength slice independently. The alternative approach, that we adopt in this work, consists in relying on a weight mask $w_{i,j}$, defined as the portion of flux that falls inside the simulated fiber aperture for each pixel $(i,j)$ in the sky plane of the CALIFA data cube.
At any given wavelength $\lambda$ the flux in fiber is thus given by
\begin{equation}
    f(\lambda)=\sum_{i,j} w_{i,j}\,f_{i,j}(\lambda).
\end{equation}

Assuming 2D Gaussian functions for the PSFs of both the fiber-fed spectroscopic observations (i.e., SDSS) and the CALIFA IFS observations, \ and neglecting any wavelength dependence of the two PSFs, the weight map is given by
\begin{equation}\label{eq:weight_map}
    w_{i,j}=\frac{1}{2\pi\sigma^2_{\mathrm{conv}}}\int_{C(x_c,y_c,D(z_s)/2)}\,dx\,dy\,e^{-\frac{(x-x_{i,j})^2+(y-y_{i,j})^2}{2\sigma^2_{\mathrm{conv}}}},
\end{equation}
where $x_{i,j}$ and $y_{i,j}$ are the coordinates of the $(i,j)$-pixel and the integral is meant to be extended to the circle $C$ centered at the center of the galaxy $(x_c,y_c)$ and having diameter $D(z_s)=3^{\prime\prime}\cdot{\mathcal{R}}_A(z_s,z_0)$.
The effective width of the Gaussian $\sigma_\mathrm{conv}$ is the one defined in Eq. \ref{eq:sigma_conv} and takes into account the native blurring of the CALIFA IFS.
As long as the PSFs can be considered independent of the wavelength (see Sect. \ref{subsub:caveats}), the weight map is computed numerically\footnote{In order to speed up the procedure, we rely on look-up tables that sample the PSF, assuming a simple 2D Gaussian function, so that the computation reduces to a simple summation of weights within a circular mask.} only once per galaxy and per simulated redshift, so substantially reducing the computational demand with respect to the direct convolution method, which would have to be repeated a number of times equal to the number of sampled wavelengths in the cubes, i.e., 1901 times.

It should be noted that, in mathematical terms, this alternative approach consists in a simple inversion on the order of the convolution integral and the integration within the fiber. In this case, we first integrate the convolution kernel over the fiber aperture to compute the weight matrix $w_{i,j}$ (equal for all wavelengths as long as the PSFs can be considered constant), and then in each wavelength slice we integrate the flux weighted by $w_{i,j}$.

We simulate the SDSS-like fiber spectra for the CALIFA sample describe in Sect. \ref{sub:CALIFA} over the redshift range $0.005<z<0.4$ (i.e., the redshift range relevant to the SDSS sample, see Sect. \ref{subsec:SDSSsample}).

\subsubsection{Illustration of the key aperture effects}\label{subsub:apersim_illustration}
\begin{figure*}
\centering
    \includegraphics[width=0.8\textwidth]{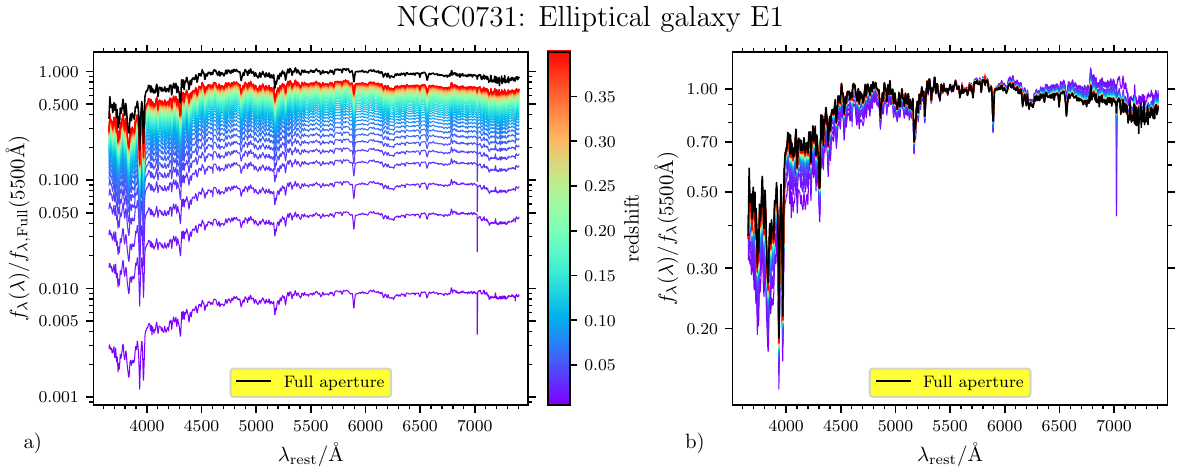}
    \includegraphics[width=0.8\textwidth]{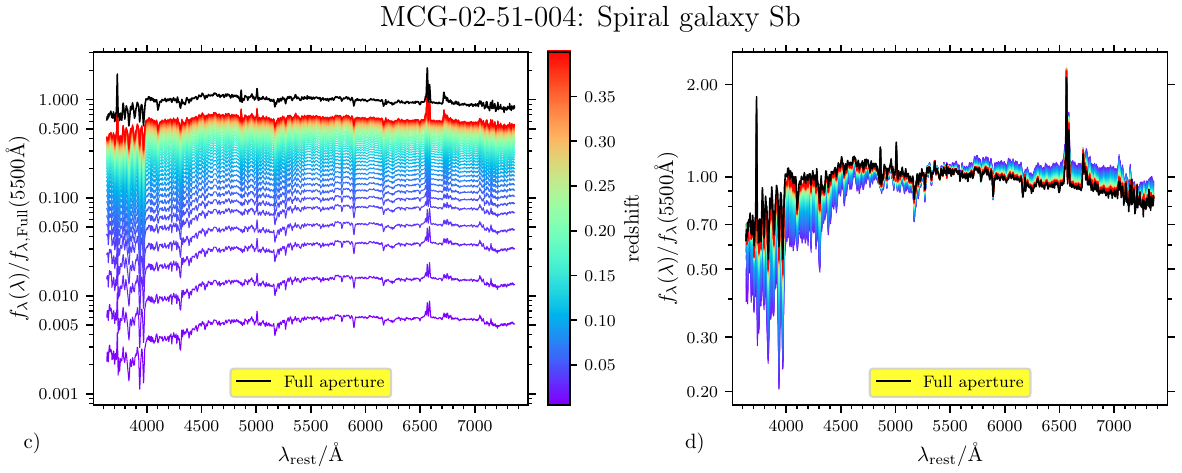}
    \includegraphics[width=0.8\textwidth]{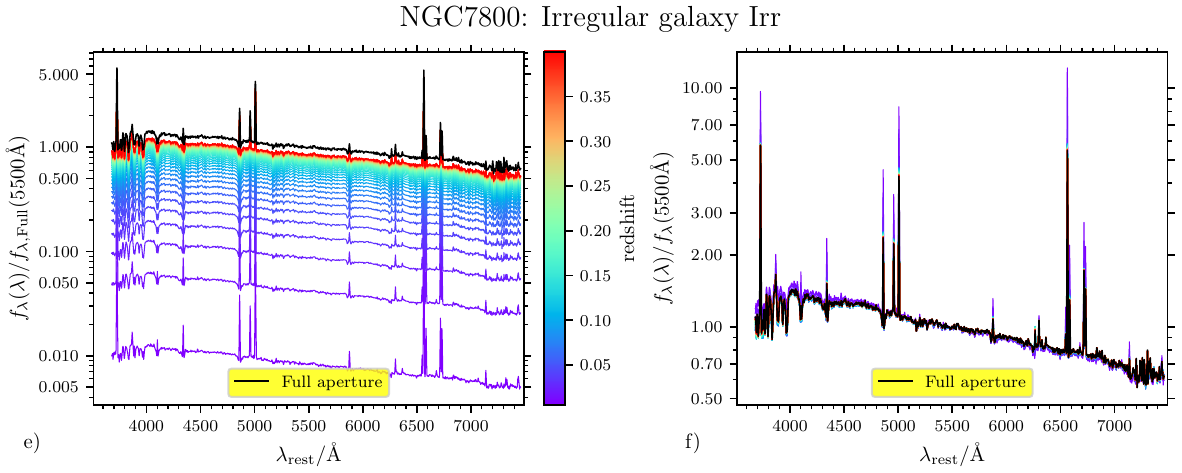}
      \caption{Spectra of three galaxies of different morphological types (elliptical, Sb spiral, and irregular) simulated to reproduce what the SDSS fiber-fed spectrograph would see over a redshift range between $0.005$ and $0.4$. We note the log scale in the $y$-axis. In all panels, the black spectrum is the one obtained from the light integrated over the full galaxy footprint of the galaxy, normalized at $5500$\,\AA. The colored spectra are the simulated fiber-spectra at the redshift indicated by the color bar. In the left panel of each galaxy (a, c, e), the fiber spectra are normalized relative to the total flux in the footprint in order to emphasize the flux loss from the fiber. In the right panels (b, d, f), all spectra are normalized to the flux density at $5500$\,\AA~in order to emphasize the changes in spectral shape and features.}\label{fig:stacked_apspec}
\end{figure*}
In Fig. \ref{fig:stacked_apspec} we illustrate the results of the fiber-aperture simulation for three galaxies representative of three broad morphological classes: 
NGC\,0731 (E1), representative of early-type galaxies (ETGs, ellipticals and S0s); MCG-02-51-004 (Sb), representative of spiral galaxies (S0a to Sd); NGC\,7800 (Irr), representative of the latest morphological types (Sdm, Sm and irregulars).  The black spectra in all panels represent the full integral over the footprint, normalized at $5500$\,\AA. The spectra plotted in color are obtained from the fiber simulations at different redshifts, according to the color bar.
In the left panels (a, c, and e for the three galaxies, resp.), their normalization is relative to the full-aperture integrated spectrum, so that the flux density at $5500$\,\AA\ corresponds to the fiber fraction. In the right panels (b, d, and f), the spectra for the fiber simulations are all normalized to unit flux density at $5500$\,\AA. We note that the spectra are plotted with a logarithmic stretch for the flux densities axis, in order to appreciate shape differences independent of the overall normalization.

From a quantitative point of view, from the left panels of Fig. \ref{fig:stacked_apspec} we note that the most dramatic variations in the integrated flux occur at $z\lesssim 0.15$, where the fraction of flux in fiber ranges between $10^{-3}$ and $0.5$. Above this critical redshift range, the simulated spectra pile up and slowly approach a limiting flux density of $\sim 0.7-0.8$ relative to the full integrated spectrum. This behavior is indicative of two regimes dominated by the two main mechanisms that regulate the fiber-aperture effects. At low $z$ the galaxies are spatially resolved and the flux collected by the fiber is mainly determined by the geometry of the fiber projection on the galaxy image. At higher $z$, the blurring by the atmospheric seeing becomes more and more relevant as the image of the galaxy asymptotically approaches the PSF. Thus, with an assumed FWHM of $1.6^{\prime\prime}$ for the seeing, even at the highest $z$ of the SDSS, the image of a galaxy is never fully included in the fiber aperture. In fact, at the limit of a point-like source, only $\sim80\%$ of the flux would be collected by the fiber, similar to what we observe for the simulated spectra at $z=0.4$.

Looking at the right panels of Fig. \ref{fig:stacked_apspec} where all spectra are equally normalized at $5500$\,\AA, we can better appreciate the changes in shape and in features (absorption and emission), i.e., the qualitative modifications that affect the estimate of the stellar population physical parameters. The strongest effects are seen for the spiral galaxy MCG-02-51-004 (panel d). The spectra simulated at $z\lesssim 0.1$ display the typical properties of an old stellar population: overall red color, a strong $4000$\,\AA~break, strong metal absorption features, and a substantial lack of emission lines. On the other hand, moving to larger $z$ the spectrum gradually transforms into the typical spectrum of a star-forming galaxy, with a blue shape, weak $4000$\,\AA~break and metal absorptions, and intense emission lines (both recombination and nebular). This phenomenology illustrates the transition from a bulge-dominated spectrum at low $z$, where the fiber collects only the central light, to a spectrum that is more representative of the whole flux, which, in this case, is dominated by the star-forming disk. Spiral galaxies, made of distinct components such as the bulge and the disk, display the most extreme variations in terms of aperture effects, in line with the expectations from the steep age (and metallicity, to a lower extent) gradients found in these galaxies (e.g., \citealt{zibetti+17,Parikh:2019}, but see also \citealt{Gonzalez-Delgado:2015aa} for somewhat quantitatively different results). Systematic, although less extreme, variations of the spectral shape and features are also seen in the template elliptical galaxy NGC\,0731: the lower-$z$ simulations display redder colors and stronger break and metal absorptions, in line with the expectations from the strongly negative metallicity gradients observed in these galaxies \citep[e.g.,][]{Zibetti:2020aa}. In the case of the irregular galaxy NGC\,7800 we observe very small variations in terms of the continuum shape, consistent with a very young stellar population at all $z$, and more evident changes in the emission lines. This can be interpreted as a consequence of a lack of systematic gradients, due to the irregular morphology and an overall young age.

In all three examples we see a substantial convergence of the normalized spectra (thus of the spectral shapes and absorptions) above $z\sim0.3$, despite the flux collected by the fiber only reaching $\sim 70\rm{\%}$ at this redshift. In fact above $z\sim0.3$ we enter the PSF-dominated regime, where the FWHM of the PSF is similar to or larger than the projected effective radius of the galaxies. The blurring is such that galaxies are seen as point-like sources and the different spectral components are mixed almost irrespective of their original spatial distribution. Thus the portion of light captured by the fiber is representative of the total light. 

For the rest of the paper we concentrate on the stellar component of the spectra.
In the two following subsections we analyze in a quantitative way the effects we just illustrated in these three examples, accounting for the full statistical diversity of the CALIFA sample.

\subsection{Estimates of the aperture biases from CALIFA data cubes: Fiber flux fractions}\label{sub:CALIFA_fluxfrac}

\begin{figure*}
    \centering
    \includegraphics[width=0.7\linewidth]{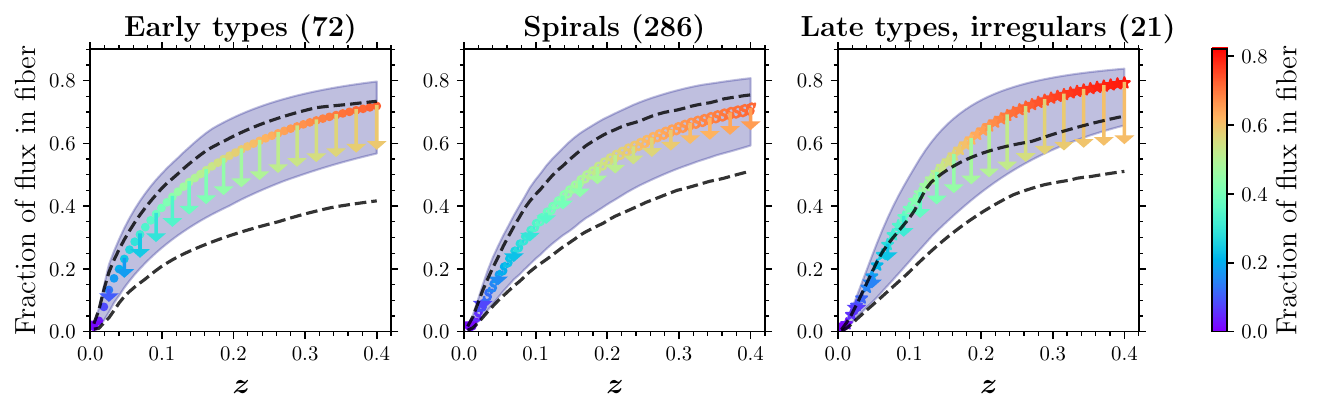}
    \caption{Trends of fractions of flux collected in fiber as a function of redshift for three broad morphological classes: ETGs (left panel), spirals (central panel), and late-type and irregular galaxies (right panel). The colored points (shaded gray areas) represent the median (16th-84th percentile range) fractions relative to the footprint (see Sect. \ref{sub:CALIFA_fluxfrac}). The vertical arrows point to the median fractions relative to the growth-curve (GC) integrated flux (i.e., the footprint fraction times the fiber fraction relative to the footprint). The dashed lines indicate the 16th and 84th percentiles for the fraction relative to the total GC flux.}
    \label{fig:fiberfracz_by_morph}
\end{figure*}

Moving from the observation of the apparent diversity of aperture effects for different morphological types, we analyze the statistical trends of the fraction of flux in fiber with redshift, separately for the three broad classes of galaxies that compose our CALIFA sample: 72 ETGs (ellipticals and S0s); 286 spiral galaxies (S0a to Sd); 21 late-type and irregular galaxies (Sdm, Sm, and irregulars). In the three panels of Fig. \ref{fig:fiberfracz_by_morph} (for the three classes, resp.) the colored points and the shaded areas display the median and the $16^{\rm{th}}-84^{\rm{th}}$ percentile range of the fraction of flux that enters the fiber, relative to the flux integrated from the CALIFA footprint (see Sect. \ref{sub:CALIFA}), as a function of simulated redshift. We also show the fractions relative to the total flux computed from the GC analysis. These fractions are lower by a factor corresponding to the footprint fraction $f_{\mathrm{footprint}}$ (eq. \ref{eq:f_fp}) and are indicated with the downward arrows for the median values of the distributions and the dashed lines for the $16^{\rm{th}}$ and $84^{\rm{th}}$ percentiles. 

From these plots we can clearly see the starkly non-linear increase of the fiber fraction as a function of redshift, common to all morphological classes: the initial steep slope at $z\lesssim0.05$ gradually decreases reaching an almost flat behavior above $z\sim 0.3$. As already pointed out in the previous section, this can be interpreted as the transition between a low-$z$ regime, where the galaxy is spatially resolved and the fiber gradually covers an increasingly larger extent of the galaxy, to a high-$z$ regime, where the galaxy is unresolved (i.e., its projected size is comparable or smaller than the PSF FWHM) and the fiber covers a constant fraction of the PSF extent, corresponding to $\sim 80 \rm{\%}$ of the flux. The different average physical size and structural properties of the galaxies in the three classes \citep{blanton03_sersic,shen_etal03} are responsible for the main differences in the shape of the fiber fraction trends. Focusing on the fractions relative to the footprint flux, we observe that extreme late-type and irregular galaxies tend to plateauing to the unresolved, high-$z$ regime more quickly than the other classes, due to their smaller intrinsic sizes. Spirals and early types exhibit very similar median trends at $z\gtrsim 0.15$, where the larger effective radius of spirals for a given luminosity is compensated by a luminosity distribution shifted to larger values for the early-type galaxies. Early types and spirals clearly differ at low $z$, where the more concentrated surface brightness profiles of the early types result in a steeper increase of the fiber fraction. As a mere consequence of the distribution in luminosity/size within each subsample, for all three classes, the scatter around the median relation is significant, on the order of $0.10-0.15$ for most of the redshift range, except for the obvious compression toward zero at the lowest $z$.

It is worth noting that considering the fiber flux fractions relative to the total GC fluxes (arrows and dashed lines) does not affect significantly the shape of the trends in general. However, while for spiral galaxies the difference with respect to the footprint-based fractions are limited to $0.1$ at most, for the other two classes systematic differences up to $0.2$ are possible. For late types or irregulars we observe also significant variations in the inter-percentile range. Such a poor correspondence between footprint and GC fluxes is indeed expected for irregular SB distributions, which are overall close to the limiting magnitude of the footprint cut.

\begin{figure*}
    \centering
    \includegraphics[width=0.8\linewidth]{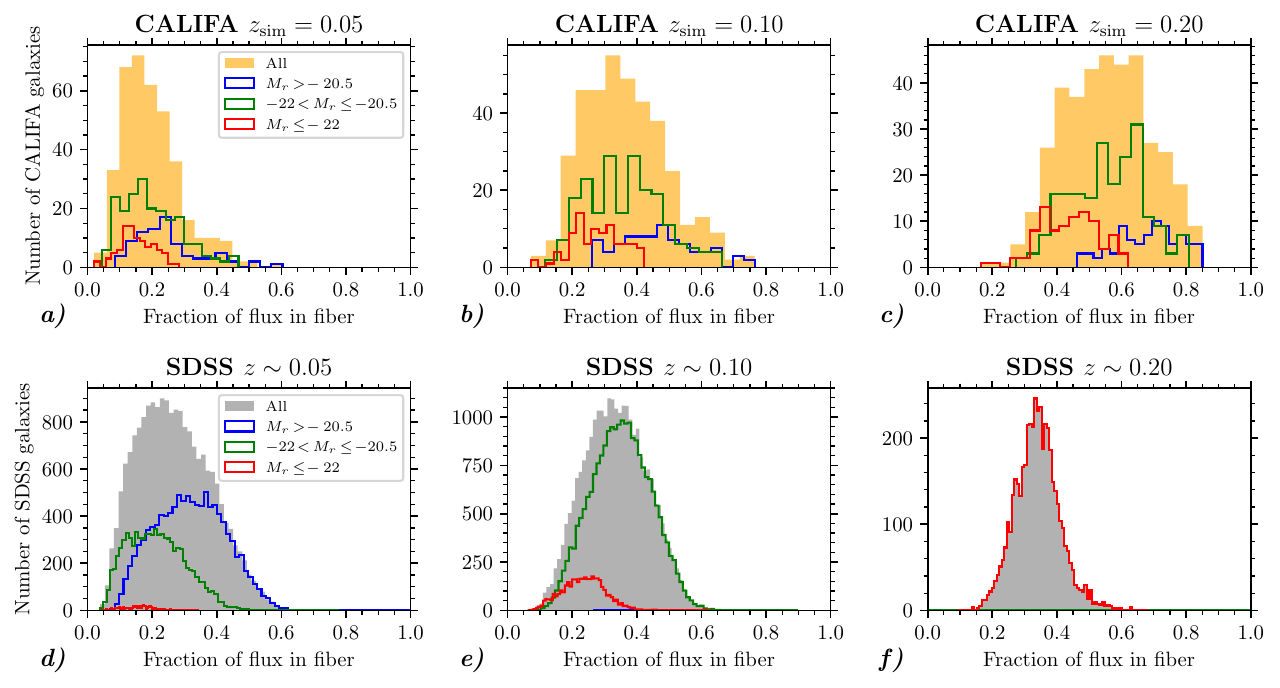}
    \caption{Distributions of the fractions of flux in fiber for the CALIFA simulated spectra (top row) and the SDSS sample (bottom row), at three redshifts: $z=0.05,~0.10,~0.20$. The orange shaded histogram represents the full CALIFA sample, while the gray-shaded histogram is the distribution for the full SDSS ($\mathrm{S/N}>10$) sample at the given redshift slice. The red, blue, and green histograms for both CALIFA and SDSS are made for three different bins of absolute magnitude, as indicated in the legend. We note that the fiber fraction for CALIFA is relative to footprint flux, while for the SDSS it is relative to the Petrosian magnitude.}
    \label{fig:fiberfrac_distr}
\end{figure*}

In Fig. \ref{fig:fiberfrac_distr} we compare the distributions of fiber flux fractions as simulated on the CALIFA sample and as measured in the SDSS sample ($\mathrm{S/N}\geq10$), at three different redshifts,\footnote{The simulations are done at the exact redshift value, while for the SDSS we consider galaxies in a thin redshift slice of $\pm 0.005$ around the reference value.} for all luminosities and in three sub-classes of luminosity. This is done in order to demonstrate the quantitative accuracy of our fiber simulations. The fractions of flux in fiber for the SDSS galaxies are obtained from the difference between the photometric ``fiber magnitudes'' (integrated from the images in a $3^{\prime\prime}$-diameter aperture) and the Petrosian magnitudes. We note that Petrosian magnitudes underestimate the total flux of a source with a virtually infinite SB profile. According to the calculations by \cite{graham_etal:2005}, the underestimate ranges between $1\%$ and $\gtrsim\!20\%$  for S\'ersic index ranging between 1 and $\gtrsim\!4$, thus approximately matching the fraction of flux missed by our IFS footprint relative to the GC total flux. Looking at the full CALIFA sample (top row, shaded golden histogram), we see the expected trend for a shift of the distribution, with the peak moving from $\sim\!0.15$ at $z_\mathrm{sim}=0.05$ to $\sim\!0.3$ at $z_\mathrm{sim}=0.1$ and to $\sim\!0.55$ at $z_\mathrm{sim}=0.2$, with a concurrent substantial broadening, in line with the results of Fig. \ref{fig:fiberfracz_by_morph}. By splitting the contributions in luminosity bins (colored histograms), we can highlight the large systematic differences among them. The most luminous galaxies are characterized by fiber fractions that are on average approximately half as large as the fiber fractions of the least luminous ones, at every redshift. The overall fiber-fraction distribution in the SDSS thus critically depends on the luminosity distribution of the sample. 

As we can see in the bottom row of Fig. \ref{fig:fiberfrac_distr}, the overall fiber-fraction distribution of the SDSS (gray-shaded histogram) displays a very mild evolution (the peak moves from $\sim\!0.25$ to $\sim\!0.35$ over the redshift range $0.05-0.2$) in contrast to the strong evolution of the distribution of the full CALIFA sample. This results from the bulk of the SDSS sample being dominated at low $z$ by low-luminosity galaxies with   higher fiber-fractions and, conversely, at high $z$ by high-luminosity galaxies with  lower fiber-fractions. In fact, if we compare the fiber-fraction distributions in the three luminosity classes in the SDSS and in the CALIFA simulations, we observe a remarkable match, especially for the intermediate luminosities. In the low-luminosity bin (only relevant for the lowest $z$), the CALIFA sample is still substantially biased toward high luminosity with respect to SDSS, thus resulting in a fiber-fraction distribution that is biased toward lower values, although broad enough to reproduce the SDSS galaxies with the largest fiber-fractions. Conversely, for the high-luminosity bin we find a more extended tail of large fiber fractions in the CALIFA simulations with respect to the SDSS. This is partly justified by the fact that the simulated fiber fractions shown in this plot are relative to the footprint flux, which represents a lower fraction of the total GC flux for higher-luminosity galaxies (see Sect. \ref{sub:CALIFA}). 

\subsection{Estimates of the aperture biases from CALIFA data cubes: Spectral shape and stellar absorption indices}\label{sub:CALIFA_aperbias}
\begin{figure*}
\centering
    \includegraphics[width=0.8\linewidth]{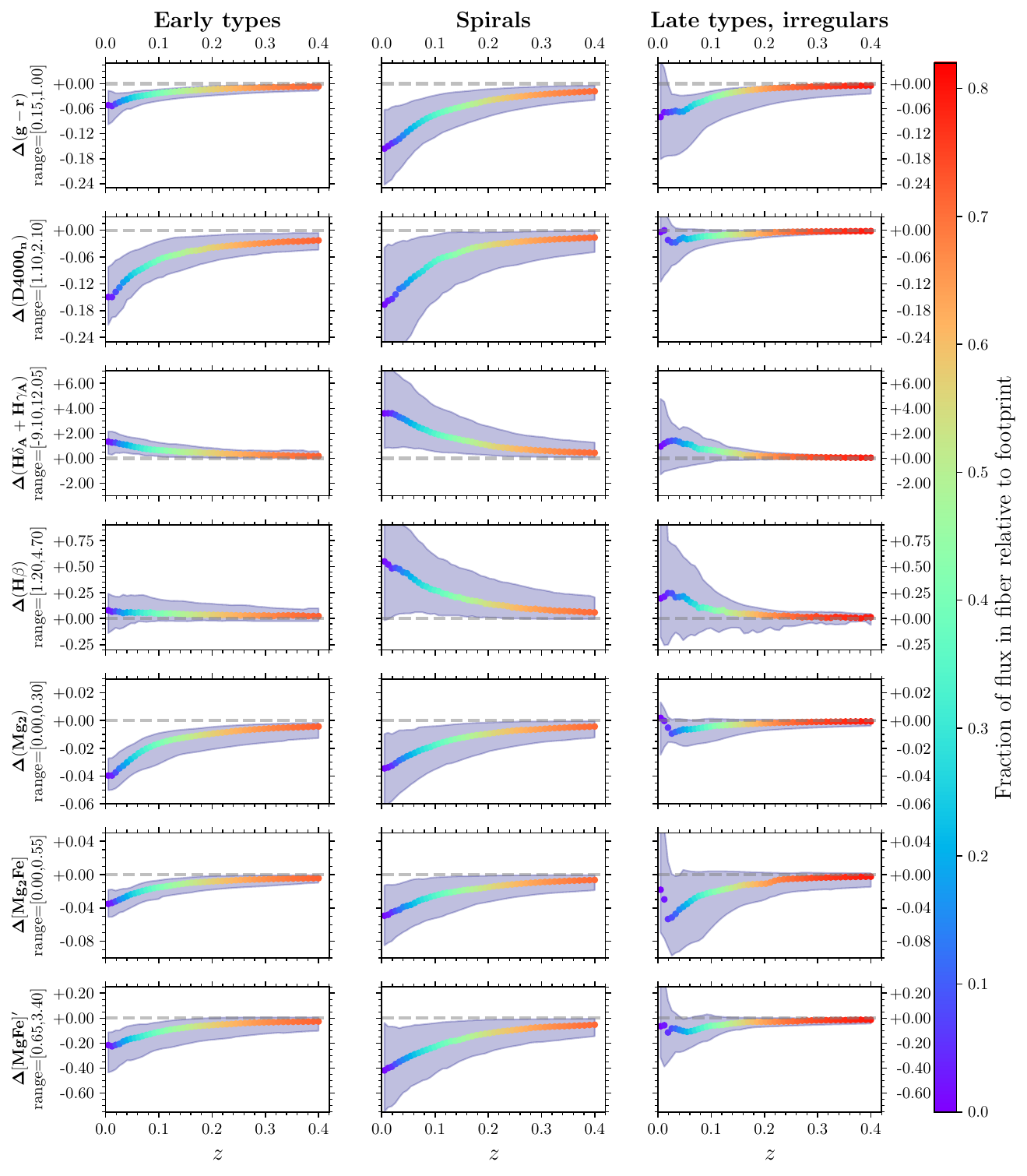}\caption{Trends of differences between indices from full integrated light and from the fiber-aperture flux (i.e., the aperture bias) as a function of redshift, for the three morphological subsamples defined in Sect. \ref{sub:CALIFA_fluxfrac}. The points indicate the median trends and are colored according to the median fraction of flux in fiber as indicated by the color bar. The gray-shaded area displays the $16\mathrm{th}-84\mathrm{th}$ percentile range, while the dashed horizontal line is the zero reference. The dynamic range of each index is given as a reference to evaluate the relevance of the aperture bias.}
    \label{fig:delta_index_morph}
\end{figure*}

In this section we illustrate the aperture biases in the intensive spectral properties (shape and absorption features) as a function of redshift, for the different morphological classes.
In Fig. \ref{fig:delta_index_morph} we present the systematic trends as a function of redshift for the differences between the values from the light within the footprint and the fiber-aperture values, $\Delta X \equiv X_\mathrm{integrated} - X_\mathrm{fiber}$, for the $g-r$ color and for a set of indices with a strong sensitivity on age and metallicity of the stellar populations: $\mathrm{D4000_n}$, $\mathrm{H\delta_A}+\mathrm{H\gamma_A}$, $\mathrm{H\beta}$ (mostly age sensitive), $\mathrm{Mg_2}$, $\mathrm{[Mg_2Fe]}$, and $\mathrm{[MgFe]^\prime}$ (mostly metallicity sensitive). As in the previous sections, the CALIFA sample is split in three morphological classes, ETGs, spirals, and late types and irregulars, represented in the three columns of the figure, respectively. The points, color-coded according to the median fraction of flux in fiber relative to the footprint, represent the median $\Delta X$ of the class as a function of $z$, while the gray-shaded areas represent the $16\mathrm{th}-84\mathrm{th}$ percentile range. The vertical scale is the same for the three classes, in order to compare the amplitude of the aperture effects for different morphologies. The full dynamical range of the color or index $X$ across all galaxies is indicated in the axis label, in order to appreciate the relevance of the aperture effect in terms of their impact on stellar population properties in a general context. 

We observe general trends that are common to all morphological classes. (i) The integrated spectra are bluer than the fiber spectra, both in $g-r$ and in $\mathrm{D4000_n}$; the strength of the Balmer absorptions is larger in the integrated spectra; on the contrary, the metal-sensitive absorption indices are weaker in the integrated spectra than in the fiber ones. These trends reflect the most common stellar population gradients observed in galaxies, i.e., negative gradients in age (mostly observed in spirals and late-type galaxies) and in metallicity (particularly strong in ETGs, but also observed in spirals and late types). (ii) The strongest aperture effects are seen at $z\lesssim 0.05$ and display a quick variation up to $z\sim\!0.2$ (corresponding to $\sim\!0.5$ of flux in fiber), above which $\Delta X$ approaches $0$ with a very mild slope as a function of $z$. By $z\!\sim\! 0.4$, the median $\Delta X$ is remarkably close to $0$ in most cases, despite the fraction of flux in fiber never exceeding $\sim 0.8$. This behavior can be promptly interpreted in the scenario introduced in Sect. \ref{subsub:apersim_illustration}, whereby we observe a transition from a spatially resolved regime at low $z$, in which trends are dictated by stellar population gradients, to the unresolved regime at higher $z$, where the light of the different galaxy components is blurred into a single unresolved source.
(iii) The largest median $\Delta X$ correspond to approximately $1/10$ up to $1/5$ of the dynamical range, thus are expected to have significant systematic effects on the inferred stellar population properties. In Sect. \ref{sec:SDSScorrected} and, more specifically, \ref{subsec:index_flows} we   show that   correcting the index values to the full galaxy-integrated values produce significant shifts in the diagnostic planes of age and metallicity.

(iv) By comparing the three different morphological classes, we observe that Spiral galaxies exhibit the largest aperture deviations in all diagnostic quantities as well as the largest galaxy-to-galaxy scatter in general. This is indeed expected as the spiral morphology is the one displaying the most evident spatial segregation of different structural components (the bulge and the disk, in particular), which are characterized by largely different stellar populations (old in the bulge, young in the disk). In the low-redshift spatially resolved regime, the transition from bulge-dominated to fully representative spectra explains the large aperture biases and the large variation as a function of $z$. On the other hand, the broad distribution of bulge-to-total ratios is responsible for the large  galaxy-to-galaxy scatter. ETGs suffer by much smaller aperture biases as far as color and Balmer indices are concerned, while the differences for $\mathrm{D4000_n}$ and the metal absorption indices are comparable with those of Spiral galaxies. The galaxy-to-galaxy scatter, however, is much smaller. This is easily understood considering the old ages that characterize ETGs across their radial extent, with only mild age gradients or even non-monotonic radial variations, and their quasi-universal steep negative metallicity gradients \citep[see, e.g.,][]{Zibetti:2020aa}. Late-type and irregular galaxies tend to saturate the fiber values to the integrated values at lower $z$ than the other two classes, due to their smaller physical size. At $z\lesssim 0.05-0.1$ (i.e., in the spatially resolved regime) they display a large scatter and, most remarkably, the median trends are irregular and non monotonic, with some reversal of the trends at the lowest redshifts, possibly due to the center being assigned to a bright young stellar cluster.

\subsubsection{Caveats}\label{subsub:caveats}
In the aperture simulation method outlined in Sect. \ref{subsub:apersim_method}, one of the assumptions is that the PSF of both the CALIFA data cubes and the SDSS spectroscopic observations is constant, and, more specifically, their wavelength dependence negligible.

In the case of CALIFA, this is fully justified by the fact
that the PSF is mainly determined by the fiber size of the PPak IFU and by the image reconstruction procedure of the dithered pointings \citep[described in][]{Sanchez:2012aa}. The final FWHM for the CALIFA PSF is assumed as $2.57^{\prime\prime}$, i.e., the median of the distribution given in \cite{Sanchez:2016aa}. Any deviation from this value is uncorrelated with galaxy type or wavelength, this is not expected to produce any systematic effect, but just contribute noise.

As for the SDSS, spectroscopic observations were carried out in relatively poor seeing conditions, with FWHM between $1.5^{\prime\prime}$ and $1.7^{\prime\prime}$ \citep{gunn+06}. 
We have run several tests simulating SDSS FWHM between 1.5$^{\prime\prime}$ and 2.5$^{\prime\prime}$.
The effect of varying the SDSS FWHM in this range is negligible in comparison to the dynamic range of the corrections (and even more so if compared to the dynamic range of the index values). In fact, differences are $\ll1/50$ of the dynamic range of the corrections for FWHM=2$^{\prime\prime}$ and approximately $1/40-1/50$ for the non-plausible value of 2.5$^{\prime\prime}$. In general we find that by underestimating the FWHM we would overestimate the corrections as a consequence of an underestimation of the blurring effect. Moreover, the bias is  stronger in galaxies with larger gradients (spirals in general, and ETGs as far as $\mathrm{D4000_n}$ and metal indices are concerned).
The small magnitude of the FWHM effect can be understood considering that the SDSS fiber is 3$^{\prime\prime}$ in diameter, hence significantly larger than the typical FWHM. In a sense, the predominant blurring is given by the fiber aperture, rather than by the PSF.
Concerning the wavelength dependence, we note that according to the Kolmogorov theory \citep[see][]{Fried:1966, Boyd:1978}, the expected dependence of the seeing FWHM is with $\lambda^{-0.2}$. This trend is also confirmed empirically on SDSS imaging by \cite{Xin:2018AJ}, although with some scatter of the power law index. The stellar absorption features we consider in this work span the wavelength range 3800 \AA~ to 5500 \AA~ (roughly). At most the expected variation of FWHM across this range is 7-8\%. If we take $\mathrm{FWHM}=1.6^{\prime\prime}$ as a reference at 5500 \AA, the FWHM at the bluest wavelength is 1.7$^{\prime\prime}$. According to our tests, such small FWHM variations result in wavelength effects that are negligible for all indices, with amplitudes $\ll1/50$ of the dynamic range.
This FWHM variation is comparable with (or marginally smaller than) the scatter of the mean FWHM across the SDSS galaxy sample, according to \cite{gunn+06}. 

In spite of the fact that, in principle, systematic effects may arise because indices at shorter wavelengths (mostly the $\mathrm{D4000_n}$ and the Balmer absorptions, chiefly age-sensitive) are slightly overcorrected with respect to those at longer wavelength (mostly Fe and Mg absorptions, chiefly metallicity-sensitive), the amplitude of such systematic effects is so small that they have no impact on stellar population parameter estimates.

Furthermore,\ we note that the aperture effects on index measures include the contribution of the velocity dispersion broadening, which can alter the index strength as measured in fixed bandpasses \citep{Worthey+1994,Worthey_Ottaviani1997ApJS}. In fact, the line-of-sight velocity distribution of the different parts of galaxy are folded into the simulated aperture spectrum and the effective broadening (``velocity dispersion'') of the spectrum is also subject to aperture effects. 
    
\section{Recipes for aperture corrections of stellar absorption indices}\label{sec:apercorr_recipes}
Relying on the fiber spectra simulated on a grid of redshifts over $0.005\!<\!z\!<\!0.4$, as described in Sect. \ref{sec:spec_simul}, we develop a procedure to correct the index values actually measured on a fiber spectrum to the values that would be measured considering the full integrated light. As pointed out in the previous sections, corrections strongly depend on galaxy morphology. However, as demonstrated by the large galaxy-to-galaxy scatter shown in Fig. \ref{fig:delta_index_morph}, a broad morphological classification is not enough to pinpoint the corrections with sufficient accuracy. In fact, using the median trends would certainly reduce the systematic bias, but would introduce a substantial scatter, well above the typical observational uncertainties. As the goal is to apply these corrections to the SDSS sample, we cannot rely on a fine morphological classification either, both because of the sample size and of the limited imaging resolution of the SDSS. After extensive testing, we decide to rely on four easily accessible and objective observable parameters, which we demonstrate to capture the variety of aperture corrections with very good accuracy. For each index we consider: the index measured in the fiber; the global $g-r$ color of the galaxy; the $r$-band physical effective radius $R_\mathrm{e}$ (in kpc); the absolute $r$-band (rest-frame) magnitude. The idea is to sample the correction function at the points corresponding to the CALIFA galaxies and apply the corrections to the SDSS galaxies by extending the function with a nearest-neighbor interpolation. Instead of working directly in a four-dimensional parameter space, we split the correction into a first-order correction, which relies on index in fiber and color alone, and a second-order correction, based on size and absolute magnitude. As we illustrate next, the first-order correction accounts for most of the systematic differences, while the second order is a refinement to attain the maximum possible accuracy.
We note that the corrections are computed for each redshift in the grid independently from the others.

\begin{figure*}
    \centering
    \includegraphics[width=0.8\linewidth]{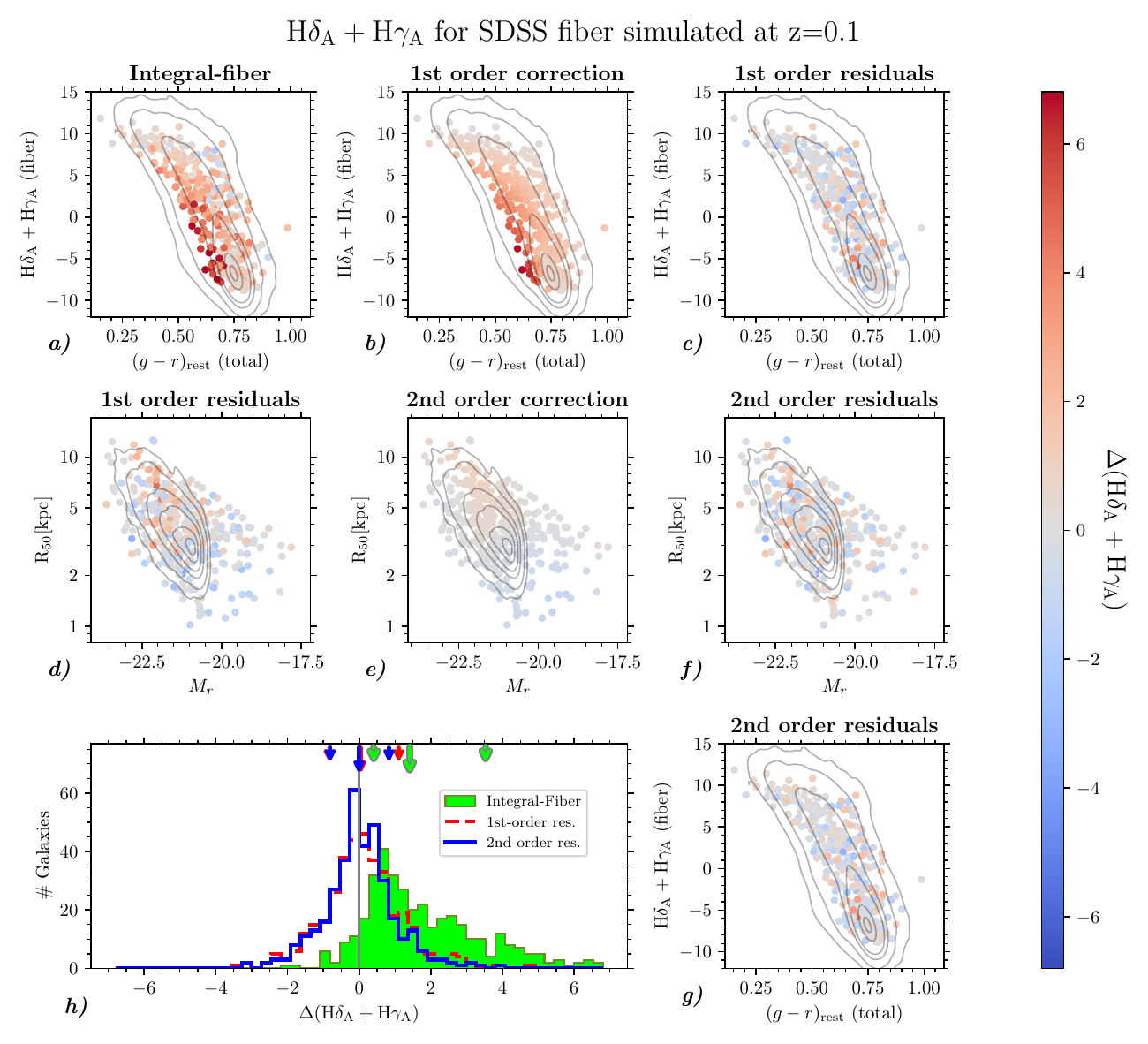}\caption{Illustration of the aperture-correction flow for the case of the $H\delta_\mathrm{A}+H\gamma_\mathrm{A}$ index at $z=0.1$. In panels a--g the filled circles represent CALIFA simulated spectra, color-coded by the difference $\Delta\left(H\delta_\mathrm{A}+H\gamma_\mathrm{A}\right)$ according to the color bar. The gray contours overlaid in each panel are the isodensity contours (enclosing 0.05, 0.20, 0.50, 0.75, 0.95, 0.99 of the sample) for SDSS galaxies with $\mathrm{S/N}\geq 10$ at $z=0.1 \pm 0.005$. Panels a--c and g display the points in the fiber-index vs. color plane, while panels d--f are in the effective radius vs. absolute magnitude plane. (a) Original difference between the index measured in the full integrated spectrum and the one in the fiber spectrum. (b) Amplitude of the first-order correction. (c) Residual difference between the index from the integrated-flux and the fiber value corrected at the first order. (d) Same as c, but in the effective radius vs. absolute magnitude plane. (e) Second-order corrections. (f) Residual difference between the index from the integrated-flux and the fiber value corrected using both the first-order and the second-order corrections. (g) Same as f, but in the fiber-index vs. color plane. The histograms in panel h display the following distributions: in filled green the original differences between integrated and fiber values (points in panel a); in dashed red the differences between integrated and fiber values corrected at the first order (points in panels c, d); in solid blue the differences between integrated and fiber values corrected at the second order (points in panels f and g). The vertical arrows on the top axis indicate the median (long arrow) and the $16\mathrm{th}$ and $84\mathrm{th}$ percentiles (short arrows) of the distributions of the corresponding color.}
    \label{fig:corr_flow_HdHg}
\end{figure*}

In the following we detail the correction procedure, which we illustrate in Fig. \ref{fig:corr_flow_HdHg} for the $H\delta_\mathrm{A}+H\gamma_\mathrm{A}$ composite index at redshift 0.1. This index is chosen due to the large amplitude of the corrections, which make all the steps very clear, but the same considerations apply qualitatively to all indices, as demonstrated by the analogous figure for the $\mathrm{[MgFe]}^\prime$ index (Fig. \ref{fig:corr_flow_MgFep}), which we present in appendix \ref{app:bias_corr_distr}. $z=0.1$ is chosen as the approximate median redshift of the SDSS. 
For a given $z$, the first step is to consider the $\Delta X$ (in this case $\Delta\left(H\delta_\mathrm{A}+H\gamma_\mathrm{A}\right)$) for each CALIFA galaxy as a function of the index measured (simulated) in fiber and the total $(g-r)_\mathrm{rest}$ color. These are shown in panel (a) as points representing individual CALIFA galaxies, color-coded by $\Delta X$ according to the color bar (the same representation is adopted also for panels b to g). We clearly observe systematic variations across this plane, which we regularize by means of a two-dimensional LOESS regression \citep{Cleveland_LOESS2D} using the Python code \textsc{loess}\footnote{Available from \url{https://pypi.org/project/loess}.} of \cite{cappellari:2013}, adopting a degree of 1 and a fraction of 0.3 (or a minimum of 20 neighboring points). The smoothed  $\Delta X$ function obtained in this way represents the first-order correction and is shown in panel (b). The residual  $\Delta X$ from this correction are shown in panel (c) in the index-in-fiber versus $g-r$ plane, and in panel (d) in the size versus absolute magnitude plane. In this latter plane systematic trends for the first-order residuals are visible. We correct for them using the same regularization approach adopted before,\footnote{The LOESS regression is applied to $\left( M_r,\log (R_{50}/\mathrm{kpc})\right)$ datapoints.} thus obtaining the second-order corrections shown in panel (e). 
Then, the second-order residuals correspond to the difference between the original $\Delta X$ values (panel a) and the sum of the first-order (panel b) and second-order corrections (panel e). Panels (f) and (g) show the residual $\Delta X$ after the two corrections in the two observed parameter planes, respectively, where we can observe the absence of significant systematic trends and a substantial consistency with random noise. 

In practice, in order to avoid the correction functions being affected by strongly deviant galaxies, after each step, we implement a $\sigma$-rejection algorithm to exclude those galaxies for which the residuals $\Delta X$ after applying the smoothed corrections exceed $3$ times the standard deviation of the full sample. The $\sigma$-rejection is applied the first time after the first calculation of the first-order corrections. The first-order corrections are then re-computed for this reduced sample and the second-order corrections are computed consequently. A $\sigma$-rejection is then applied again based on the second-order residuals. Finally, the full set of corrections is re-computed for the sample surviving the two $\sigma$-rejections. Typically, no more than a handful of galaxies are rejected per index and per redshift value.

In panel (h) we plot the distributions of the differences of the corrected index values with respect to the integrated values, only for the galaxies surviving the $\sigma$-rejections (red and blue histograms). For reference, the green histogram shows the original differences for the values measured in fiber, which display an evident bias of $\sim 1.4$\,\AA~ (median, long vertical green arrow) and a significant skewness toward positive difference, with a tail up to $7$\,\AA. After correcting the fiber values by adding the first-order corrections and the second-order corrections, we obtain the red-dashed histogram and blue histogram, respectively. As we can see from the medians of the distributions (marked by the long arrows, in red and blue respectively), the bias becomes negligible already after the first correction. The dispersion (represented by the 16th and 84th percentiles of the distributions marked with the short arrows) is also substantially reduced, already with the first-order correction. Nonetheless, the second-order correction is particularly important to reduce the tail and the skewness of the distribution, as apparent comparing the position of the 84th percentile with the second-order correction and with the first-order correction only.

In table \ref{tab:deltaX_distr} in appendix \ref{app:bias_corr_distr} we report the values of the median and percentiles of the $\Delta X$, without corrections and after the first- and second-order corrections, for the six stellar absorption indices that are also represented in Fig. \ref{fig:delta_index_morph}, for three representative redshifts: $0.05$, $0.1$, and $0.2$. As in the example of $H\delta_\mathrm{A}+H\gamma_\mathrm{A}$, in all cases most of the bias is removed by the first-order corrections, while the second-order correction mainly impact the dispersion and the skewness of the distributions. Remarkably, in our CALIFA simulations of SDSS observations, the scatter in the residuals after the correction procedure is smaller than the typical measurement errors in the SDSS, indicating that uncertainties in the aperture corrections have a negligible impact on the final error budget of the indices.

The points displayed in panel (b) and (e) of Fig. \ref{fig:corr_flow_HdHg} represent the discrete sampling, restricted to the CALIFA sample, of the first- and second-order corrections functions for the index $X$, $\mathit{Corr_1\left((g-r)_\mathrm{rest,total},\mathrm{X_{fiber}}\right)}$ and
$\mathit{Corr_2\left(M_r,R_{50}/\mathrm{kpc}\right)}$, respectively, for a given redshift $z$. In order to extend these corrections function over a continuous domain, we need to define rules to interpolate them at any possible coordinate. We opt for the simplest solution of a nearest neighbor interpolation, based on observationally motivated metrics in the 2d parameter spaces. For the two planes we define the distances between any given observed point and the points corresponding to the $i$-th CALIFA galaxy  respectively as
\begin{equation}\label{eq:d1}
d_{1,i} = \sqrt{\frac{\left((g-r)_\mathrm{obs}-(g-r)_i\right)^2}{\sigma_{g-r}^2}+\frac{\left(X_\mathrm{obs}-X_i\right)^2}{\sigma_X^2}}
\end{equation}
and
\begin{equation}\label{eq:d2}
d_{2,i} = \sqrt{\frac{\left(M_{r,\mathrm{obs}}-M_{r,i}\right)^2}{\sigma_{M_r}^2}+\frac{\log^2 \left(R_{50,\mathrm{obs}}/R_{50,i}\right)}{\sigma^2_{\log R_{50}}}},
\end{equation}
where the various $\sigma$ are suitably defined observational errors (see Sect. \ref{subsec:SDSSsample}), which effectively define the metric distances. Once the indices $i_{1,\mathrm{min}}$ and $i_{2,\mathrm{min}}$ are determined as those minimizing $d_1$ and $d_2$, respectively, the corrections are given by:  $\mathit{Corr_1}=\mathit{Corr_1\left((g-r)_{i_{1,\mathrm{min}}},\mathrm{X_{fiber}}_{i_{1,\mathrm{min}}}\right)}$ and
$\mathit{Corr_2}=\mathit{Corr_2\left(M_{r,i_{2,\mathrm{min}}},R_{50,i_{2,\mathrm{min}}}/\mathrm{kpc}\right)}$. We note that in general $i_{1,\mathrm{min}} \neq i_{2,\mathrm{min}}$, or, in other words, the first- and second-order corrections are not necessarily derived from the same CALIFA galaxy.

It should be noted that this approach based on the error metric allows one to best use the available information for each galaxy. For instance, in the case of the first-order corrections, the index in fiber can be more or less constrained, depending on the spectral S/N. The metric defined above gives it accordingly more or less weight relative to the $g-r$ color.
Since the corrections for the CALIFA galaxies are computed for a discrete grid of $z$, for a given $z_\mathrm{obs}$ the closest simulated redshift on the grid shall be used.

\section{Correcting the stellar absorption indices of the SDSS sample against the aperture bias}\label{sec:SDSScorrected}
\subsection{The SDSS sample: Description and estimates of the aperture corrections}\label{subsec:SDSSsample}
In this section we focus on a spectroscopic sample of SDSS galaxies drawn from the seventh data release (DR7) and specifically suited for stellar population analysis. The fully detailed description of the catalogue sources and sample selection is provided in \cite{Mattolini:2025}. Here we summarize the main properties and provide the essential references. The DR7 \citep{SDSS_DR7} is the final data release of the SDSS-II, which completed the single 3-arcsec-fiber observations of about $\sim\!930\,000$ galaxies in the local Universe initiated by the SDSS \citep{SDSS,EDR}.
The primary source of our dataset are the MPA-JHU catalogues,\footnote{The MPA-JHU catalogues are available at the following link: \url{https://wwwmpa.mpa-garching.mpg.de/SDSS/DR7/raw_data.html}.} which collect $927~552$ photometric and spectroscopic measurements of galaxies (including repeated spectroscopic observations of the same photometric counterparts). 
These catalogues are matched to the NYU Value-Added Galaxy Catalog \citep[][NYU-VAGC hereafter]{2005AJ....129.2562B}, from which the photometric information (observed- and rest-frame) is derived.
After rejecting low-quality spectroscopic measurements ($\sim 5\%$ of the sample), multiple spectroscopic measurements of unique galaxies were combined as error-weighted means. These include redshift $z$, stellar velocity dispersion $\sigma_\star$, and a full set of commonly adopted stellar absorption indices, including the $\mathrm{D4000_n}$ break and the Lick-system indices. The indices are directly measured on the observed spectra at the native resolution and without correcting for $\sigma_\star$, after removing the nebular emission lines as fitted by single Gaussians \citep[see][]{brinchmann+04}.
The master catalog obtained in this way collects unique observations for $825~263$ SDSS-DR7 galaxies.

The aperture corrections for the indices are computed following the recipes provided in Sect. \ref{sec:apercorr_recipes}, based on the observational parameters obtained as follows.
The $r$-band absolute magnitudes $M_r$ and the rest-frame $(g-r)$ colors are derived from the Petrosian absolute magnitudes, k-corrected to $z=0$, provided by the NYU-VAGC,\footnote{Specifically, available at this link: \url{https://sdss.physics.nyu.edu/vagc-dr7/vagc2/kcorrect/kcorrect.none.petro.z0.00.fits}.} and further corrected to our adopted standard cosmology. The half-light radii $R_{50}$ are computed from the Petrosian half-light radii (in $r$-band) as measured by the standard SDSS photo pipeline and included in the MPA-JHU catalog gal\_fnal.\footnote{\url{https://wwwmpa.mpa-garching.mpg.de/SDSS/DR7/Data/gal_fnal_dr7_v5_2.fit.gz}} These angular sizes are converted to physical units of kpc, based on our adopted cosmology. As for the errors that define the metrics of Eq. \ref{eq:d1} and \ref{eq:d2}, we start by considering the formal errors provided by the catalogs. In order to regularize the metrics and avoid excessive distortions of the parameter planes, as well as to avoid unrealistically low uncertainties that give too much weight to any parameter, we enforce lower limits to the errors: 0.05 mag for the $g-r$ color, 0.1 mag for $M_r$, and 0.1'' for the angular size of $R_{50}$. We note that these limits result in typical errors corresponding to approximately 1/10 of the dynamical range for the $g-r$ color (to be compared with a similar error-to-dynamical range ratio along the index axis, in the first-order correction plane), and to roughly 1/50 of the dynamical range for $M_r$ and for $\log R_{50}$ axes (in the second-order correction plane).

In the next two sections we discuss the main systematic effects of applying the aperture corrections to the measures of absorption index of the SDSS-DR7 galaxy sample. Following \cite{Mattolini:2025} we consider only galaxies that are part of the Main Galaxy Sample (MGS), hence with $14.5\,\text{mag}\leq r_\mathrm{petro,~dered}\leq 17.77\,\text{mag}$ \citep{strauss_etal02} in the redshift range $0.005<z\leq0.22$. We also discard galaxies with poor or problematic photometry and velocity dispersion exceeding $375\,\kms$ to exclude possible non-physical measurements (possibly large errors).
Results are showcased for two different subsamples: a subsample of $354~977$ galaxies with $\mathrm{S/N}\geq10$, which we use for age-sensitive indices ($\mathrm{D4000_n}$ and Balmer line indices), and a subsample of $89~852$ galaxies with $\mathrm{S/N}\geq20$, which we use for metallicity-sensitive indices. We note that these samples are not complete in mass nor in luminosity, thus the population analysis presented in the following sections is just meant to provide a comparison of the relative corrections of various indices across the stellar population parameter space.

\subsection{Systematic effects of aperture corrections on absorption indices}\label{subsec:index_flows}
In this section we analyze the effects of the aperture corrections by looking at the distribution of SDSS galaxies in popular index-index diagnostic diagrams.

\subsubsection{Galaxy distribution in the ``Balmer plane''}
The $\mathrm{H\delta_A}+\mathrm{H\gamma_A}$ versus $\mathrm{D4000_n}$ plane (the ``Balmer plane'', hereafter) is a powerful diagnostic of a galaxy's SFH and of its stellar age in particular \citep[see, e.g.,][]{kauffmann+03b}. SSPs exhibit bell-shaped evolutionary tracks as a function of age on this plane. The youngest ages correspond to the lowest values of  $\mathrm{D4000_n}$ and intermediate Balmer absorption strength. As age increases, the $\mathrm{D4000_n}$ increases, while the strength of the Balmer indices initially increases, until they reach a maximum of $\sim 20$\,\AA~for ages of a few hundred million years \citep[model-dependent; see, e.g., ][]{Mattolini:2025}. After the peak, SSPs move across the ``Balmer plane'' by steadily increasing $\mathrm{D4000_n}$ and decreasing the Balmer strength. As shown by \cite{kauffmann+03b}, continuous SFHs over a long time span occupy a stretched sequence in this plane, which stays substantially below the peak of the bell-shaped curve of the SSPs at young ages, and merges with the old part of the SSPs at old ages. As opposed, bursty SFHs tend to occupy the regions below the bell-shaped curve and scatter about the sequence of continuous SFHs. As it turns out, the vast majority of SDSS galaxies follow the continuous SFH sequence, although with some scatter, which can be only partly accounted for by the observational errors. 

\begin{figure*}
    \centering
    \includegraphics[width=0.8\linewidth]{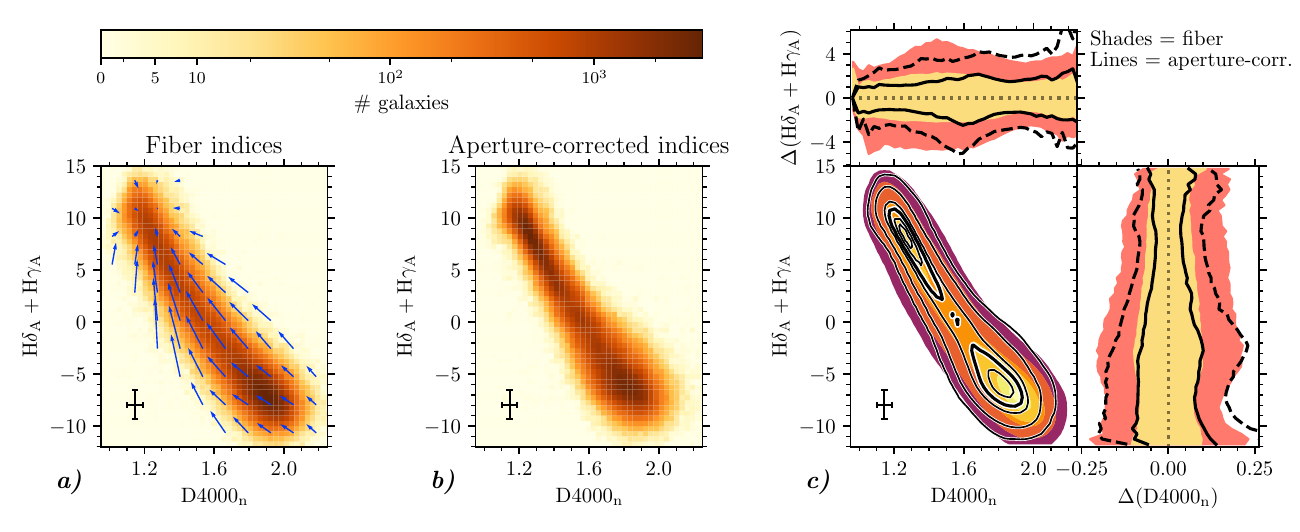}\caption{Impact of the aperture corrections of the indices on the Balmer plane $\mathrm{H\delta_A}+\mathrm{H\gamma_A}$ vs. $\mathrm{D4000_n}$, for the SDSS sample of galaxies with $\mathrm{S/N} \geq 10$. Panels a and b display the distributions of the galaxies before (a) and after (b) aperture corrections. The blue arrows in panel a indicate the median shift of galaxies due to the aperture corrections at different locations of the plane. The median error bars are also indicated. In panel c contour levels (shade-filled levels and line contours, for the original fiber values and for the aperture corrected values, respectively) are overlaid in order to facilitate the comparison of the two distributions. The levels correspond to quantiles at 0.05, 0.20, 0.50, 0.75, 0.95, and 0.99 of the sample distribution. The two side-panels of panel c illustrate the scatter of the distributions about the median relations. In the right-hand side panel, the 2.5, 16, 84, and 97.5 percentiles of the distribution of $\mathrm{D4000_n}$ about its median, as a function of $\mathrm{H\delta_A}+\mathrm{H\gamma_A}$, are represented by the shaded areas and by the different lines, for the fiber indices and for the corrected indices, respectively. The top-side panel represents the distributions of  $\mathrm{H\delta_A}+\mathrm{H\gamma_A}$ about its median, as a function of $\mathrm{D4000_n}$.}
    \label{fig:corrections_balmer_plane}
\end{figure*}

Panel (a) of Fig. \ref{fig:corrections_balmer_plane} displays the distribution of the $\mathrm{S/N}\geq 10$ sample in the Balmer plane, based on the indices as measured in the SDSS fiber spectra, with typical errors reported in the error-bar in the bottom left corner. As expected, the galaxies distribute along the continuous-SFH sequence already identified by \cite{kauffmann+03b}. In panel (b) we can see the analogous distribution obtained with the aperture-corrected indices. It is readily apparent that in the corrected distribution (i) the sequence is remarkably narrower and (ii) the density of galaxies is suppressed in the bottom-right corner (corresponding to old stellar populations) and, conversely, is enhanced going toward the top-left end of the sequence (corresponding to younger stellar populations). The reason for these changes is easily understood by looking at the blue arrows over-plotted to the original index distribution. These arrows represent the median corrections that are applied to galaxies in a neighborhood of the base point of the vector, thus indicating the average flow of displacement of galaxies in the plane. The arrows indicate that galaxies in general move along the sequence, from the old corner toward the young one. There is also evidence for the flow to converge onto the sequence in the upper-left part, which explains why the sequence gets narrower in this part. It is noteworthy that a hint of hook-like structure in the distribution appears in the upper left corner, indicating the presence of galaxies with stellar populations with ages younger than the turnaround age of the SSPs.

The most significant change, which is also the most relevant one for the demographics of galaxies' stellar population, is the emergence of a strong bimodality in the distribution. While the distribution of the original fiber indices displayed a main peak of old stellar population and a long tail with a minor and shallow peak at young ages, in the aperture-corrected distribution the main peak is located in the young region and has an elongated shape, while the old peak is now less prominent. Most notably, the two peaks are now separated by a very clear saddle. This is caused by the galaxies with largest displacement due to aperture corrections being those with the intermediate value of the indices, hence intermediate ages. These are typically spiral galaxies, which display the largest stellar population gradients, especially going from the bulge to the disk, consistently with what we pointed out in Sect. \ref{sub:CALIFA_aperbias} and Fig. \ref{fig:fiberfracz_by_morph}. In this way, the middle part of the age sequence is depleted, enhancing the younger peak.

In order to emphasize and quantify the changes due to the aperture corrections in the distributions of Fig. \ref{fig:corrections_balmer_plane} (panels a and b), in the main plot of panel (c) we show the discretized distribution of the original fiber indices (different color shades) and overplot the density contours of the aperture-corrected ones. Each level includes 0.99, 0.95, 0.75, 0.50, 0.20, and 0.05 of the sample, from the outermost/darkest to the innermost/lightest one, respectively. From this plot one can see that, after applying aperture corrections, the 5\% of the galaxies at the highest density live in the young peak, while the 5\% top density was in the old peak before correcting. Notably, while the young peak does not move, despite its substantial enhancement, the old one moves up and left along the sequence. The right-side plot of panel (c) illustrates the distribution of galaxies about the sequence, by showing the percentiles of the distribution about the median  $\mathrm{D4000_n}$ (represented by the dotted line at $0$) as a function of $\mathrm{H\delta_A}+\mathrm{H\gamma_A}$. The yellow shaded area and the solid black lines represent the 16th--84th percentile range for the aperture-corrected indices and the fiber indices, respectively; the pink dashed area and the dashed lines are the same for the 2.5th--97.5th percentiles. Similarly, the top plot displays the same information for the distribution of $\mathrm{H\delta_A}+\mathrm{H\gamma_A}$ about the median as a function of $\mathrm{D4000_n}$. As we can observe from this plot, the implementation of the aperture corrections does not alter the width of the old part of the distribution (which is already significantly broader than the young one in the original fiber indices), but it reduces  the width of the distribution for the young galaxies by roughly a factor of 2.

Interestingly, this may indicate that galaxies follow tighter index-index relations (hence, they have even more correlated stellar population properties) than it appears from raw uncorrected indices. In fact, what we observe prior to aperture corrections are the distributions convolved with the distribution of aperture biases (and of observational errors). By applying aperture corrections we effectively deconvolve the raw distribution, thus obtaining a sharper view of the real underlying galaxy distribution.

\subsubsection{Galaxy distribution in the  \texorpdfstring{$\mathrm{H\delta_A}+\mathrm{H\gamma_A}$ versus $\mathrm{[MgFe]}^\prime$}{HdeltaA+HgammaA vs. [MgFe]'} plane}
\begin{figure*}
    \centering
    \includegraphics[width=0.8\linewidth]{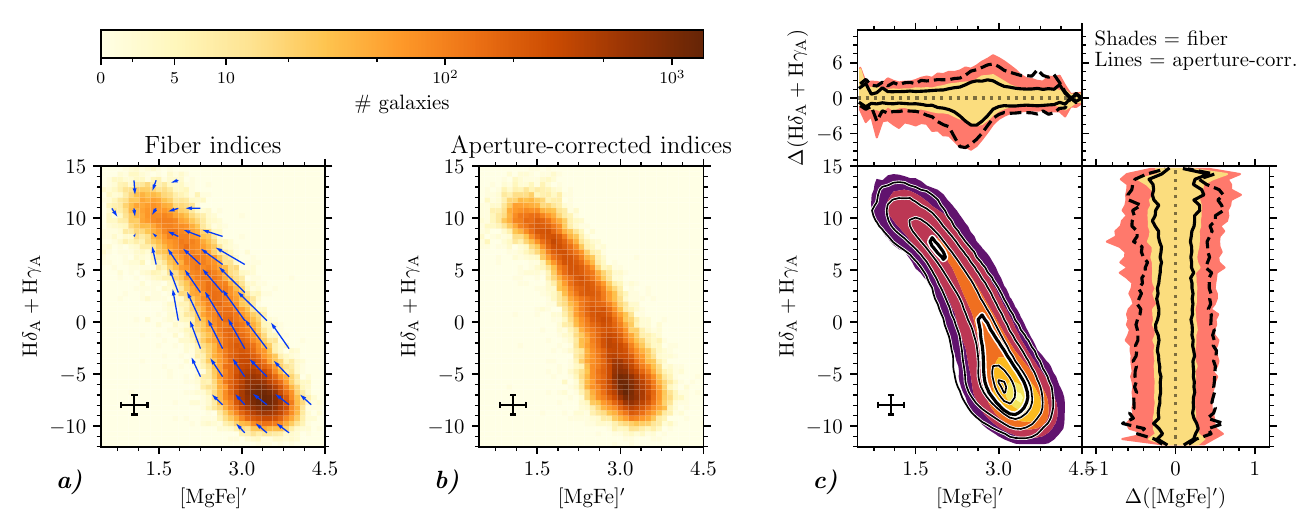}\caption{Impact of the aperture corrections of the indices on the metallicity-sensitive plane $\mathrm{H\delta_A}+\mathrm{H\gamma_A}$ vs. $\mathrm{[MgFe]^\prime}$, for the SDSS sample of galaxies with $\mathrm{S/N} \geq 20$. Same as Fig. \ref{fig:corrections_balmer_plane}, but with $\mathrm{[MgFe]^\prime}$ instead of $\mathrm{D4000_n}$.}
    \label{fig:corrections_MgFep_plane}
\end{figure*}

In the same way as we analyzed the distribution of galaxies and the variations due to aperture corrections in the Balmer plane, in Fig. \ref{fig:corrections_MgFep_plane} (analogous to Fig. \ref{fig:corrections_balmer_plane}) we study how aperture corrections affect the distribution in the metallicity diagnostic plane of $\mathrm{H\delta_A}+\mathrm{H\gamma_A}$ versus $[\mathrm{MgFe}]^\prime$, for the galaxy sample with $\mathrm{S/N}\geq 20$. Also in this case, we observe that galaxies distribute along a relatively narrow sequence, in which $\mathrm{H\delta_A}+\mathrm{H\gamma_A}$ is anti-correlated with $[\mathrm{MgFe}]^\prime$. The highest density of galaxies (neglecting statistical corrections) occurs at the lowest values of $\mathrm{H\delta_A}+\mathrm{H\gamma_A}$ and the highest values of $[\mathrm{MgFe}]^\prime$, corresponding to old and metal-rich stellar populations. Contrary to the Balmer plane, in this case there is no apparent bimodality, even after applying aperture corrections, which is connected to the fact that the distribution fades toward low $[\mathrm{MgFe}]^\prime$. The amplitude of the shift in $[\mathrm{MgFe}]^\prime$ is approximately uniform across the sequence, except around the youngest part. As a consequence, the aperture-corrected distribution is overall shifted by $\sim -0.1$ in $[\mathrm{MgFe}]^\prime$ at fixed Balmer absorption strength, thus implying an overall decrease of the metallicity. This is consistent with the expectations based on the ubiquitous negative metallicity gradients \citep[e.g.,][]{Zibetti:2020aa,Parikh:2019,Gonzalez-Delgado:2014ab}.
Noteworthy, the aperture corrections lead to a mild reduction of the scatter about the sequence in its young/metal-poor extent, although much less conspicuous than in the Balmer plane.

\subsection{Impact of aperture corrections on empirical scaling relations}
\begin{figure*}
    \centering
    \includegraphics[width=0.8\linewidth]{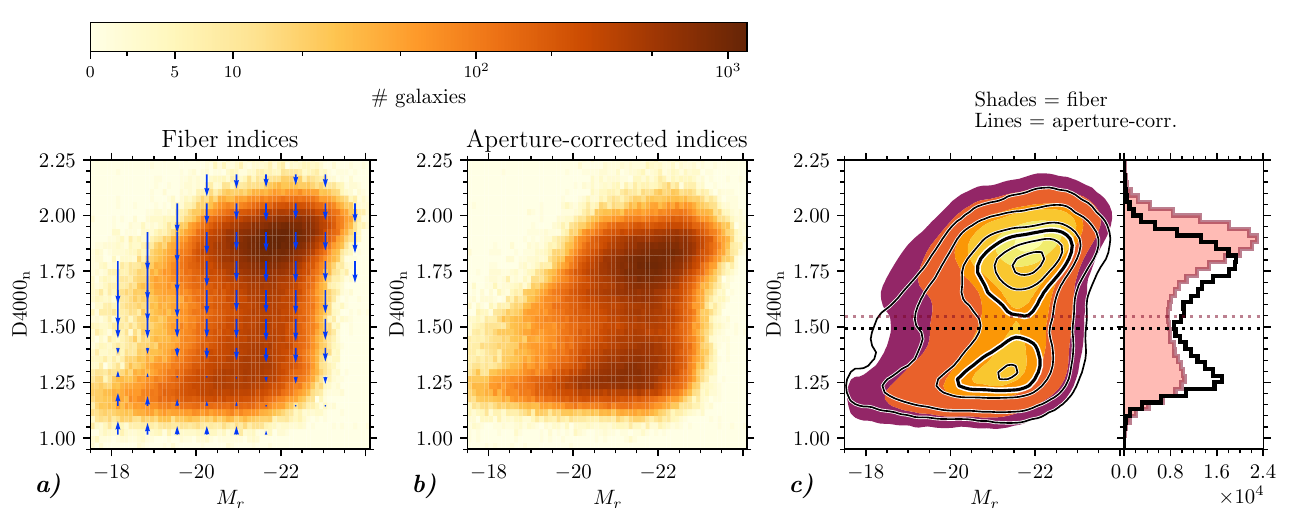}\caption{Impact of the aperture corrections of the $\mathrm{D4000_n}$ index on the bimodality-diagnostic plane $\mathrm{D4000_n}$ vs. absolute magnitude $M_r$, for the SDSS sample with  $\mathrm{S/N} \geq 10$. Panels (a) and (b) and the main plot of panel (c) are analogous to Figs. \ref{fig:corrections_balmer_plane} and \ref{fig:corrections_MgFep_plane}. The side-panel of panel (c) displays the projected distributions in $\mathrm{D4000_n}$ for the original fiber $\mathrm{D4000_n}$ (filled pink and dark red histogram) and for the aperture-corrected $\mathrm{D4000_n}$ (black empty histogram). The dotted horizontal lines mark the position of the minima of the two distributions, respectively.}
    \label{fig:corrections_mag_D4000n_bimodality}
\end{figure*}

As observed in the previous paragraphs, aperture corrections affect the bimodal distribution of galaxies in stellar population properties. The empirical bimodality observed in the distribution of $\mathrm{D4000_n}$, as a function of luminosity or stellar mass \citep[e.g.,][]{kauffmann+03b}, across different environments and cosmic time \citep[e.g.,][]{2017A&A...605A...4H,WuP-F+2018}, is one of the most basic, yet most powerful diagnostic of how galaxies evolve along different pathways, resulting in different SFHs. In Fig. \ref{fig:corrections_mag_D4000n_bimodality} we illustrate how the distribution (and its bimodality, in particular) of $\mathrm{S/N}\geq 10$ galaxies in the absolute magnitude $M_r$ versus $\mathrm{D4000_n}$ plane changes because of the aperture corrections. Analogously to figures \ref{fig:corrections_balmer_plane} and \ref{fig:corrections_MgFep_plane}, panel (a) shows the distribution using the original fiber $\mathrm{D4000_n}$ index, while in panel (b) the aperture-corrected indices are used. The vertical blue arrows display the median displacement of galaxies located in bins centered at the base-point of the arrow, due to the aperture corrections (which only apply to $\mathrm{D4000_n}$, hence their vertical direction). The sequence of young stellar populations at low $\mathrm{D4000_n}$ is not significantly affected by the aperture corrections. With only a few exceptions of very small amplitude, the galaxies in the rest of the plane are corrected to smaller values. The largest negative corrections apply to the galaxies in the ``green valley'', in between the blue young sequence (low $\mathrm{D4000_n}$) and the red old sequence (high $\mathrm{D4000_n}$). In panel (c) we can directly and quantitatively compare the distributions before and after aperture corrections. The level contours and shading indicate isodensity contours enclosing different percentages of galaxies (as in Fig. \ref{fig:corrections_balmer_plane} and \ref{fig:corrections_MgFep_plane}), while the histograms represent the distribution in $\mathrm{D4000_n}$, for the original fiber indices (filled pink histogram) and for the corrected ones (empty black histogram), respectively.  (i) The young peak of the distribution is substantially enhanced in its amplitude, yet without changing its location significantly. Conversely, (ii) the old peak is depleted and (iii) moves to lower $\mathrm{D4000_n}$ values by $\sim0.15$. The minimum of the distribution in $\mathrm{D4000_n}$ is marked with dotted lines: pink for the uncorrected indices and black for the corrected ones. The minimum can be used as an empirical separation between galaxies dominated by young and old stellar populations, respectively. Aperture corrections lower the separation line by some 0.05.

\begin{figure}
    \centering
    \includegraphics[width=0.9\linewidth]{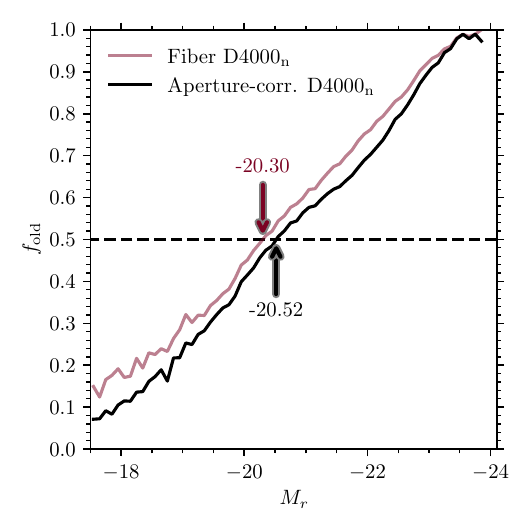}
    \caption{Fraction of old galaxies $f_\mathrm{old}$ as a function of absolute magnitude $M_r$ for the SDSS galaxy sample at $\mathrm{S/N} \geq 10$. Galaxies are classified as old based on whether they lie above the dividing line corresponding to the minimum of the $\mathrm{D4000_n}$ distribution (Fig. \ref{fig:corrections_mag_D4000n_bimodality}c), making this classification dependent on aperture corrections. Trends based on uncorrected values (dark red line) and aperture-corrected values (black line) are shown. The vertical arrows denote the transition magnitudes for each dataset, marking the luminosity above which old galaxies predominate.}
    \label{fig:old_fract_transition}
\end{figure}

As an example of the relevance of aperture corrections to accurately characterize galaxy populations in terms of their stellar properties, we focus on the well established transition that occurs from the dominance of young stellar populations at low luminosities to the dominance of old stellar populations at high luminosities \citep[e.g.,][]{kauffmann+03b, gallazzi+05}. In Fig. \ref{fig:old_fract_transition} we plot the fraction of old galaxies $f_\mathrm{old}$, i.e., those having $\mathrm{D4000_n}$ larger than the separation defined above,\footnote{The separation value is defined independently in the two cases as the minimum of the corresponding $\mathrm{D4000_n}$ distribution.} against the absolute magnitude $M_r$, in black for the aperture-corrected measurements and in dark red for the uncorrected measurements. At any fixed magnitude, the old fraction is reduced by a few percentage points up to approximately 10\% because of aperture corrections. This implies a shift of the transition magnitude at which the $f_\mathrm{old}=50\%$ to higher luminosities by 0.22 mag. These numbers are just meant to be illustrative on the order of magnitude of the effects of the aperture corrections. A dedicated analysis based on physical parameters (mean stellar age and stellar mass) and including proper statistical weights is presented in \cite{Mattolini:2025}.

\section{Summary and concluding remarks}\label{sec:conclusions}
In this paper we  presented a quantitative assessment of the amplitude of fiber-aperture effects on the measurements of stellar absorption indices of galaxies in the nearby Universe as obtained by the Legacy SDSS. We leveraged the simultaneous spatial coverage and spatial resolution of the CALIFA survey to simulate the spectra that the SDSS fiber-fed spectrograph would collect at different redshifts, also taking into account  typical seeing effects. Taking advantage of the representativeness of CALIFA over a large span of luminosities and morphologies, we studied the systematic trends of the index corrections, i.e., the differences between the indices measured by integrating over the full galaxy extent and the ones integrated within the fiber. We find the following (see Sect. \ref{sub:CALIFA_aperbias} and Fig. \ref{fig:delta_index_morph}): (i) Typical corrections for the most commonly adopted indices are $\gtrsim15\%$ of the dynamical range for $z\sim 0.02$, roughly halving at $z\sim 0.1$ (the median $z$ of the SDSS) and slowly approaching a negligible level above $z\sim 0.2$. At $z\sim 0.4$ all corrections are negligible, due to the reduced angular size of galaxies and the smearing effect of the seeing. (ii) Spiral galaxies display the largest aperture effects, due to their strong internal gradients, which is linked to their structural differentiation, while ETGs exhibit a reduced effect in the Balmer lines, due to their overall lack of young stellar populations. Late-type and irregular galaxies display small systematic effects, yet the object-to-object variation is large. Therefore, correcting the aperture effects for these galaxies individually is virtually impossible based on a statistical approach, yet statistically we expect these corrections to be negligible on average.

By studying the correlations between the aperture corrections and different observable parameters, we were able to devise a procedure to estimate the correction to apply to any absorption index for a given galaxy at a given redshift (Sect. \ref{sec:apercorr_recipes} and Fig. \ref{fig:corr_flow_HdHg}). We split the corrections into first-order and second-order corrections. The first-order corrections can be estimated from the index as observed in the fiber and the global $g-r$ rest-frame color. The second-order corrections are estimated based on the absolute $r$-band magnitude $M_r$ and the physical half-light radius $R_{50}$ (in kpc) of the galaxy. Statistically, most of the bias in the index measurement with respect to the full integrated value is corrected at the first order. The second-order corrections mainly affect the scatter and become more  negligible with decreasing  signal-to-noise ratio of the observations. As shown in Table \ref{tab:deltaX_distr}, the scatter in the second-order residuals is always smaller than the typical measurement error in the SDSS.

We applied these correction recipes to the SDSS-DR7 dataset and analyzed their impact on the distribution of galaxies in popular index-index diagnostic planes. The general trends agree with the expectations based on the well-known correlations between stellar populations and their gradients and morphology. In particular, (i) we obtain an enhancement of the bimodality in the distribution in the age-sensitive Balmer plane ($\mathrm{H\delta_A}+\mathrm{H\gamma_A}$ vs. $\mathrm{D4000_n}$) as a consequence of a substantial flow of galaxies of intermediate age toward younger ages. (ii) In the metallicity-sensitive $\mathrm{H\delta_A}+\mathrm{H\gamma_A}$ versus $[\mathrm{MgFe}]^\prime$ plane, there is a general trend to weaker metal absorption, which can be interpreted as direct consequence of ubiquitous negative metallicity gradients.
Galaxies display evident correlations in these two planes, which are already well defined in the original indices from fibers. (iii) Aperture corrections tighten these correlations significantly, by reducing the scatter about the median relations by approximately a factor of 2 in the young--metal-poor regime. This is indeed a strong indication of the goodness of our corrections. They appear to be able to reduce the scatter connected to the random index variations due to the diverse redshift distribution of galaxies, and reveal a picture in which a galaxy's stellar population properties are tightly linked to one another. In Appendix \ref{app:link_structure} we also report evidence that aperture-corrected indices correlate significantly better with structural parameters, such as the concentration index ($R_{90}/R_{50}$) and the absolute magnitude $M_r$, than uncorrected indices do.

Finally, we illustrated the potential impact of aperture corrections on the quantitative assessment of galaxy bimodality, based on their distribution in the $\mathrm{D4000_n}$ versus $M_r$ plane. We showed a systematic overestimate of the fraction of old galaxies by a few percentage points up to 10\% (at $M_r\sim 22$), which, in turn, results in an underestimate of the transition luminosity (i.e., the luminosity at which the number of old galaxies equals the number of young ones) by more than 0.2 mag.
We note that these numbers do not take the statistical completeness of the sample into account, but are just based on raw galaxy number counts. A thorough analysis of the stellar population properties and their scaling relations based on the aperture-corrected indices and fully accounting for statistical corrections is presented in \cite{Mattolini:2025}.

The full set of the data required for the aperture corrections as well as the tables with the aperture-corrected indices for the SDSS-DR7 sample are made available on public repositories\footnote{\url{https://www.basta.inaf.it},\ and VizieR \url{https://vizier.cds.unistra.fr}}. This database will serve as a fundamental benchmark for models of galaxy evolution and as a zero point to consistently trace the evolution of galaxy stellar populations at higher redshift, based on surveys such as WEAVE-StePS \citep{Iovino+2023} and 4MOST-StePS \citep{4MOST-StePS} at $z\sim 0.55$, LEGA-C (\citealt{LEGA-C}; see, e.g., Gallazzi et al. 2025a submitted, \citealt{Gallazzi_paperII:2025}) at $z\sim 0.7$, MOONRISE \citep{MOONRISE_Maiolino2020} at $1\lesssim z\lesssim 2.5$, and studies at even higher $z$ with JWST such as JWST-SUSPENSE \citep{SUSPENCE_Slob2024}. We also note that this methodology can be straightforwardly applied to correct aperture effects in any finite-aperture (fiber or slit) spectroscopic survey, provided that a reference sample of galaxies observed in IFS is available.

\begin{acknowledgements}
      We thank the referee for insightful comments that have improved the clarity of the paper and highlighted significant aspects of the works.
      SZ acknowledges support from the INAF-Minigrant-2023 "Enabling the study of galaxy evolution through unresolved stellar population analysis" 1.05.23.04.01. ARG acknowledges support from the INAF-Minigrant-2022 "LEGA-C" 1.05.12.04.01. DM acknowledges support from the Italian national inter-university PhD programme in Space Science and Technology. LSD, ARG and SZ  acknowledge support from PRIN-MUR project “PROMETEUS” financed by the European Union - Next Generation EU, Mission 4 Component 1 CUP B53D23004750006.
\end{acknowledgements}

%
%
\bibliographystyle{aa} 

\bibliography{aa57018-25}

\begin{appendix} 
\section{Aperture bias and residual distributions for a set of popular indices}\label{app:bias_corr_distr}
In this appendix we report the statistics for the distributions of the differences $\Delta X$ (opposite of the bias) between the aperture-free integrated index $X$ and the values of the index as measured in the fiber (no correction), the fiber index corrected at the first order, and the fiber index corrected at the second order (i.e., first- and second-order corrections applied). This is presented for the six popular absorption indices presented as illustration in the main text: $\mathrm{D4000_n}$, $\mathrm{H\delta_A}+\mathrm{H\gamma_A}$, $\mathrm{H\beta}$, $\mathrm{Mg_2}$, $\mathrm{[Mg_2Fe]}$, and $\mathrm{[MgFe]^\prime}$. In Table \ref{tab:deltaX_distr} we report the median and the $16\mathrm{th}$ and $84\mathrm{th}$ percentiles of the distributions, computed at three representative redshifts: $z=0.1$ as the median redshift of the SDSS, and $z=0.05$ and $z=0.2$ as representative of the low- and high-redshift regime, respectively. These distributions are computed based on the simulations presented in Sect. \ref{sec:spec_simul} of the main text.

In the absence of any correction, we note significant median biases, which decrease from low to high redshift. The first-order corrections in all cases remove the median bias (nearly) completely, while the additional implementation of the second-order corrections reduce the scatter and remove the skewness in the distributions. In the header line, besides each index name we report the median error in the SDSS, for $\mathrm{S/N}\geq 10$ for $\mathrm{D4000_n}$, $\mathrm{H\delta_A}+\mathrm{H\gamma_A}$, $\mathrm{H\beta}$, and for $\mathrm{S/N}\geq 20$ for $\mathrm{Mg_2}$, $\mathrm{[Mg_2Fe]}$, and $\mathrm{[MgFe]^\prime}$. The scatter in the second-order residuals is in all cases equal to or less than the median errors in the SDSS data.

\begin{sidewaystable}
\caption{Statistics of differences between total integrated indices, aperture indices, and aperture-corrected indices}    \label{tab:deltaX_distr}
    \centering
     \begin{tabular}{p{0.08\textwidth}|p{0.12\textwidth}|p{0.12\textwidth}|p{0.12\textwidth}|p{0.12\textwidth}|p{0.12\textwidth}|p{0.12\textwidth}}
         \hline \hline
         \makebox[\linewidth][c]{\centerline{\small Correction}}&
         \makebox[\linewidth][c]{\centerline{\small $\mathrm{D4000_n}~(\langle\epsilon\rangle=0.05)$}} & 
         \makebox[\linewidth][c]{\centerline{\small  $\mathrm{H\delta_A}+\mathrm{H\gamma_A}~(\langle\epsilon\rangle=1.39)$}} &
         \makebox[\linewidth][c]{\centerline{\small  $\mathrm{H\beta}~(\langle\epsilon\rangle=0.53)$}} &
         \makebox[\linewidth][c]{\centerline{\small  $\mathrm{Mg_2}~(\langle\epsilon\rangle=0.011)$}} &
         \makebox[\linewidth][c]{\centerline{\small  $\mathrm{[Mg_2Fe]}~(\langle\epsilon\rangle=0.019)$}} &
         \makebox[\linewidth][c]{\centerline{\small  $\mathrm{[MgFe]^\prime}~(\langle\epsilon\rangle=0.236)$}}\\
         \hline \hline
         \noalign{\vspace{3pt}}
         \multicolumn{7}{c}{$z=0.05$}\\
         \hline
         {\small None     } & \makebox[\linewidth][c]{\centerline{\small $-0.11~[{-0.23},{-0.03}]$}} & \makebox[\linewidth][c]{\centerline{\small $+2.05~[{+0.64},{+5.32}]$}} & \makebox[\linewidth][c]{\centerline{\small $+0.29~[{+0.01},{+0.75}]$}} & \makebox[\linewidth][c]{\centerline{\small $-0.025~[{-0.046},{-0.008}]$}} & \makebox[\linewidth][c]{\centerline{\small $-0.032~[{-0.062},{-0.009}]$}} & \makebox[\linewidth][c]{\centerline{\small $-0.248~[{-0.534},{-0.063}]$}} \\
{\small 1st order} & \makebox[\linewidth][c]{\centerline{\small $-0.00~[{-0.06},{+0.06}]$}} & \makebox[\linewidth][c]{\centerline{\small $+0.00~[{-1.10},{+1.47}]$}} & \makebox[\linewidth][c]{\centerline{\small $+0.01~[{-0.24},{+0.24}]$}} & \makebox[\linewidth][c]{\centerline{\small $+0.001~[{-0.012},{+0.011}]$}} & \makebox[\linewidth][c]{\centerline{\small $+0.000~[{-0.022},{+0.018}]$}} & \makebox[\linewidth][c]{\centerline{\small $+0.015~[{-0.182},{+0.158}]$}} \\
{\small 2nd order} & \makebox[\linewidth][c]{\centerline{\small $-0.00~[{-0.05},{+0.05}]$}} & \makebox[\linewidth][c]{\centerline{\small $+0.03~[{-1.10},{+1.18}]$}} & \makebox[\linewidth][c]{\centerline{\small $+0.01~[{-0.20},{+0.23}]$}} & \makebox[\linewidth][c]{\centerline{\small $-0.000~[{-0.011},{+0.010}]$}} & \makebox[\linewidth][c]{\centerline{\small $+0.000~[{-0.020},{+0.017}]$}} & \makebox[\linewidth][c]{\centerline{\small $+0.007~[{-0.161},{+0.147}]$}} \\

         \hline
         \noalign{\vspace{3pt}}
         \multicolumn{7}{c}{$z=0.10$}\\
         \hline
         {\small None     } & \makebox[\linewidth][c]{\centerline{\small $-0.06~[{-0.14},{-0.01}]$}} & \makebox[\linewidth][c]{\centerline{\small $+1.41~[{+0.41},{+3.53}]$}} & \makebox[\linewidth][c]{\centerline{\small $+0.19~[{+0.00},{+0.57}]$}} & \makebox[\linewidth][c]{\centerline{\small $-0.017~[{-0.034},{-0.004}]$}} & \makebox[\linewidth][c]{\centerline{\small $-0.022~[{-0.045},{-0.004}]$}} & \makebox[\linewidth][c]{\centerline{\small $-0.175~[{-0.407},{-0.034}]$}} \\
{\small 1st order} & \makebox[\linewidth][c]{\centerline{\small $-0.00~[{-0.04},{+0.04}]$}} & \makebox[\linewidth][c]{\centerline{\small $+0.05~[{-0.84},{+1.11}]$}} & \makebox[\linewidth][c]{\centerline{\small $-0.00~[{-0.18},{+0.21}]$}} & \makebox[\linewidth][c]{\centerline{\small $+0.001~[{-0.010},{+0.009}]$}} & \makebox[\linewidth][c]{\centerline{\small $+0.002~[{-0.017},{+0.015}]$}} & \makebox[\linewidth][c]{\centerline{\small $+0.005~[{-0.151},{+0.135}]$}} \\
{\small 2nd order} & \makebox[\linewidth][c]{\centerline{\small $-0.00~[{-0.04},{+0.04}]$}} & \makebox[\linewidth][c]{\centerline{\small $-0.01~[{-0.81},{+0.83}]$}} & \makebox[\linewidth][c]{\centerline{\small $+0.00~[{-0.15},{+0.18}]$}} & \makebox[\linewidth][c]{\centerline{\small $-0.000~[{-0.008},{+0.008}]$}} & \makebox[\linewidth][c]{\centerline{\small $+0.000~[{-0.016},{+0.013}]$}} & \makebox[\linewidth][c]{\centerline{\small $-0.000~[{-0.136},{+0.116}]$}} \\
         \hline
         \noalign{\vspace{3pt}}
         \multicolumn{7}{c}{$z=0.20$}\\
         \hline
         {\small None     } & \makebox[\linewidth][c]{\centerline{\small $-0.03~[{-0.08},{-0.00}]$}} & \makebox[\linewidth][c]{\centerline{\small $+0.70~[{+0.14},{+1.89}]$}} & \makebox[\linewidth][c]{\centerline{\small $+0.10~[{-0.01},{+0.30}]$}} & \makebox[\linewidth][c]{\centerline{\small $-0.009~[{-0.020},{-0.001}]$}} & \makebox[\linewidth][c]{\centerline{\small $-0.011~[{-0.028},{-0.001}]$}} & \makebox[\linewidth][c]{\centerline{\small $-0.095~[{-0.251},{-0.009}]$}} \\
{\small 1st order} & \makebox[\linewidth][c]{\centerline{\small $+0.00~[{-0.03},{+0.03}]$}} & \makebox[\linewidth][c]{\centerline{\small $+0.01~[{-0.56},{+0.70}]$}} & \makebox[\linewidth][c]{\centerline{\small $-0.01~[{-0.12},{+0.16}]$}} & \makebox[\linewidth][c]{\centerline{\small $+0.001~[{-0.007},{+0.007}]$}} & \makebox[\linewidth][c]{\centerline{\small $+0.001~[{-0.012},{+0.011}]$}} & \makebox[\linewidth][c]{\centerline{\small $+0.002~[{-0.101},{+0.096}]$}} \\
{\small 2nd order} & \makebox[\linewidth][c]{\centerline{\small $-0.00~[{-0.02},{+0.02}]$}} & \makebox[\linewidth][c]{\centerline{\small $-0.01~[{-0.60},{+0.55}]$}} & \makebox[\linewidth][c]{\centerline{\small $-0.00~[{-0.12},{+0.12}]$}} & \makebox[\linewidth][c]{\centerline{\small $-0.000~[{-0.005},{+0.005}]$}} & \makebox[\linewidth][c]{\centerline{\small $-0.000~[{-0.011},{+0.010}]$}} & \makebox[\linewidth][c]{\centerline{\small $+0.007~[{-0.089},{+0.081}]$}} \\

\hline \hline
         \noalign{\vspace{10pt}}

    \end{tabular}
    \tablefoot{Median and percentiles of the distributions of the differences $\Delta X$ between the aperture-free integrated index $X$ and the values of the index as measured in the fiber (correction = ``None''), the fiber index corrected at the first order (correction = ``1st order'', and the fiber index corrected at the second order (i.e., first- and second-order corrections applied, correction = ``2nd order''), for six absorption indices (columns 2--7) and three sets of simulations at $z=0.05,\,0.1,\,0.2$. In each column we report the median along with the $16\mathrm{th}$ and $84\mathrm{th}$ percentiles in square brackets.}
\end{sidewaystable}

\begin{figure*}
    \centering
    \includegraphics[width=0.8\linewidth]{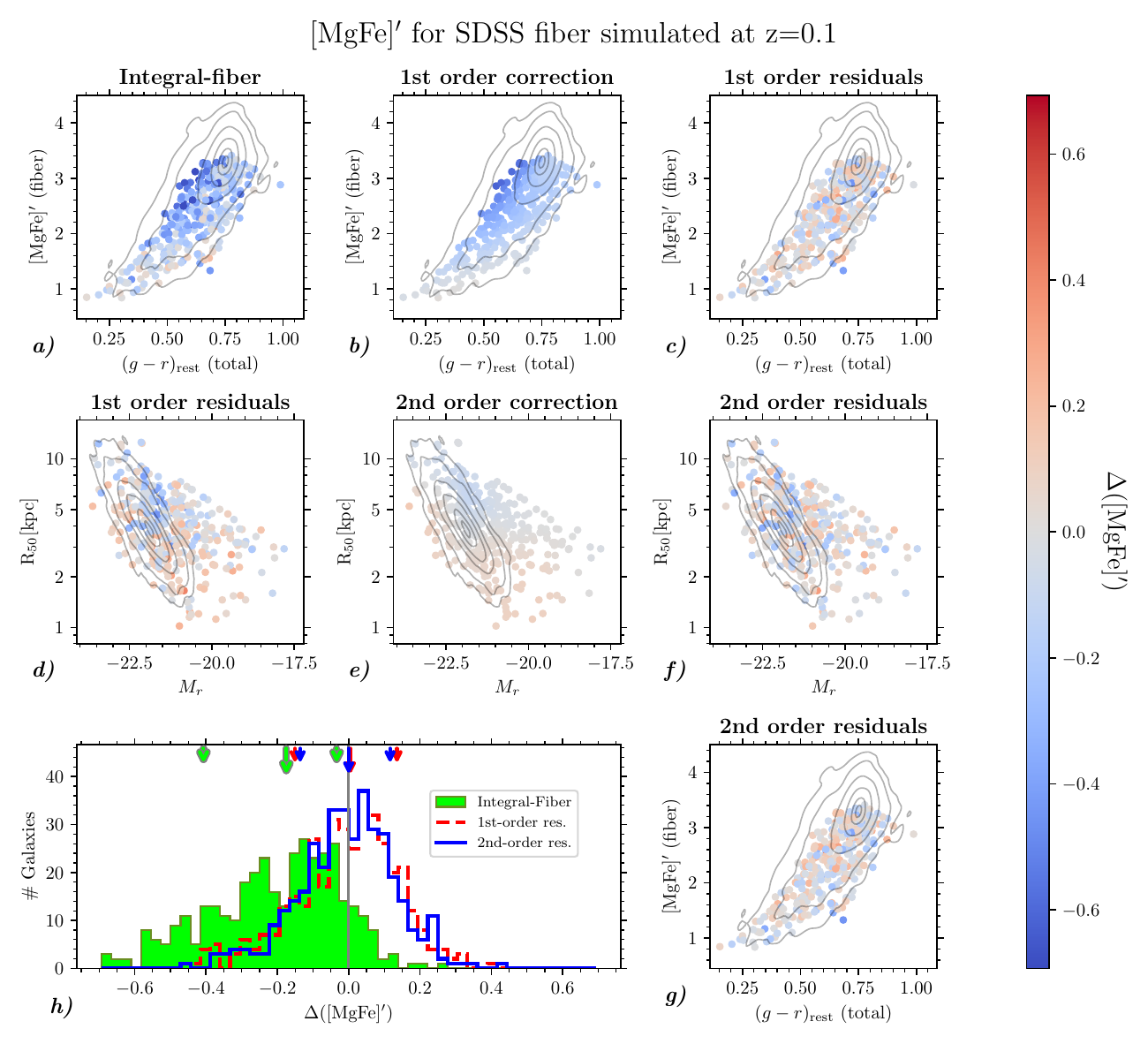}\caption{Illustration of the aperture-correction flow for the case of the $\mathrm{[MgFe]}^\prime$ index at $z=0.1$ (see caption to Fig. \ref{fig:corr_flow_HdHg} for a full description of the plots). The isodensity contours in this case are relative to the SDSS galaxies with $\mathrm{S/N}\geq 20$.}
    \label{fig:corr_flow_MgFep}
\end{figure*}

In order to complement the illustration of the aperture-correction computation, we include the analogous of Fig. \ref{fig:corr_flow_HdHg} for the metallicity-sensitive index $\mathrm{[MgFe]}^\prime$ as Fig. \ref{fig:corr_flow_MgFep}. Except for the reversal of the sign of the corrections, very similar qualitative considerations apply.

\section{A tighter relation between stellar populations and galaxy structure}\label{app:link_structure}
\begin{figure}[htbp]
    \centering
    \includegraphics[width=\linewidth]{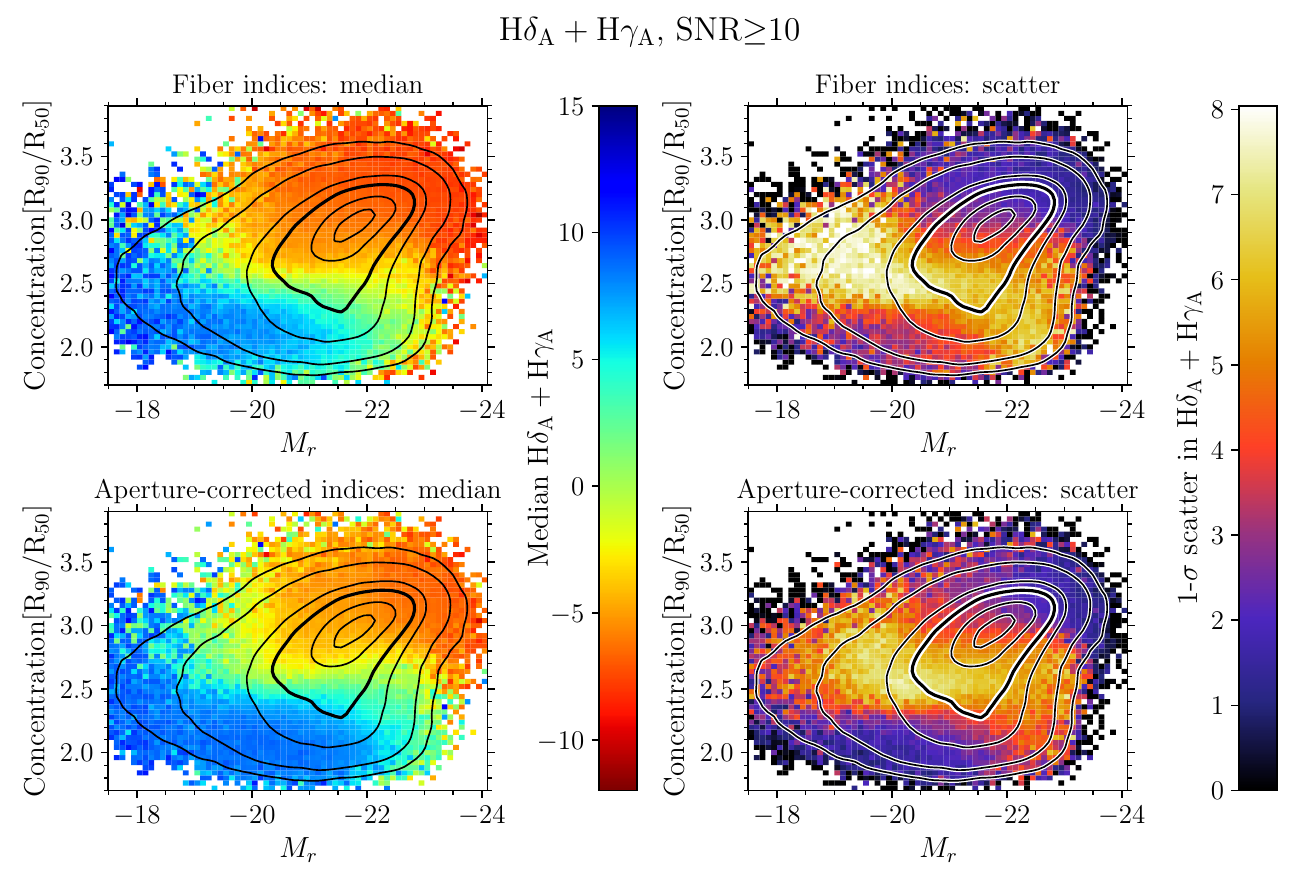}\caption{Maps of the median $\mathrm{H\delta_A}+\mathrm{H\gamma_A}$ and its scatter (computed as one-half of the $16\mathrm{th}$--$84\mathrm{th}$ percentile range) for SDSS-DR7 galaxies with $\mathrm{S/N}\geq 10$ in 2D bins in the plane of concentration vs. absolute $r$-band magnitude $M_r$. The maps in the top row are obtained with the original fiber indices, while those in the bottom row are computed with aperture-corrected indices. Iso-density contours are overlaid, including fractions of 0.05, 0.20, 0.50 (thick contour), 0.75, 0.95, and 0.99 of the sample, respectively.}
    \label{fig:mag_conc_HdHg}
\end{figure}

\begin{figure}[ht]
    \centering
    \includegraphics[width=\linewidth]{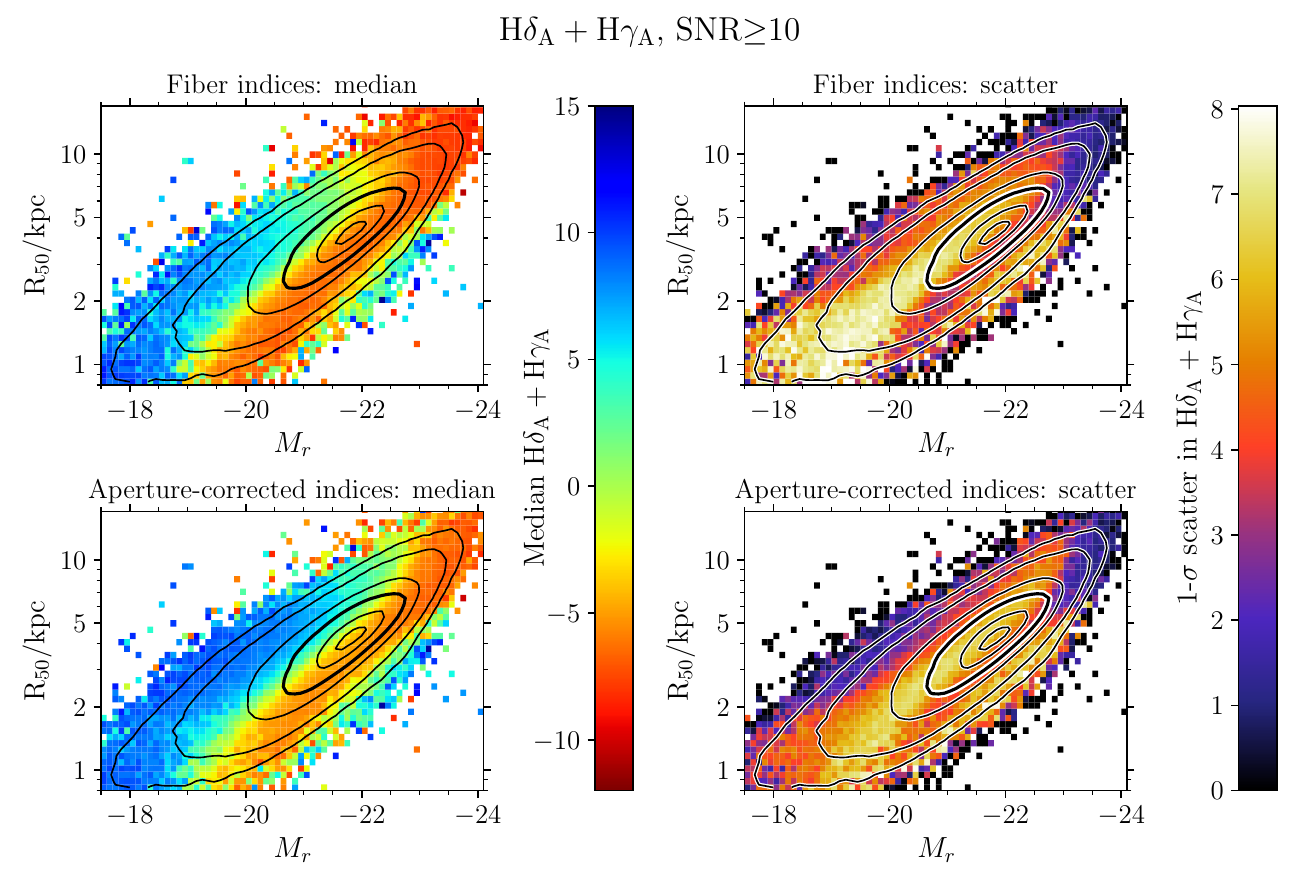}\caption{Same as Fig. \ref{fig:mag_conc_HdHg}, but for the  plane of half-light radius $R_{50}$ vs. absolute $r$-band magnitude $M_r$.}
    \label{fig:mag_R50_HdHg}
\end{figure}

The stellar populations in galaxies are known to be linked to the structure of the galaxy itself \citep[e.g.,][]{scott+2017,zibetti_gallazzi2022}. To explore these correlations we map the median value and the width of the distributions of stellar population parameters across planes of different structural parameters, for the SDSS-DR7 galaxy sample with $\mathrm{S/N}\geq 10$. In figures \ref{fig:mag_conc_HdHg} and \ref{fig:mag_R50_HdHg} we consider the Balmer composite index  $\mathrm{H\delta_A}+\mathrm{H\gamma_A}$ as a tracer of stellar population properties in the planes defined by the concentration index $\mathrm{R_{90}/R_{50}}$ versus the absolute $r$-band magnitude $M_r$ and by the half-light radius $\mathrm{R_{50}}$ versus $M_r$, respectively. 

The left panels show the map of the median $\mathrm{H\delta_A}+\mathrm{H\gamma_A}$ of the galaxies in 2D parameter space bins, while the right panels show the corresponding 1-$\sigma$ dispersion in $\mathrm{H\delta_A}+\mathrm{H\gamma_A}$, given by half of the $16\mathrm{th}$--$84\mathrm{th}$ percentile range. Top rows use uncorrected fiber indices; bottom rows employ aperture-corrected values (second order). Density contours illustrate the sample distribution. The most striking effect of the aperture corrections is a significant decrease of the scatter in $\mathrm{H\delta_A}+\mathrm{H\gamma_A}$ at fixed structural parameters, which is particularly notable at faint magnitudes and low concentrations and/or large sizes (for the given magnitude). The median maps are mainly affected in terms of a systematic shift. However, interestingly in the size-magnitude plane, the implementation of aperture corrections results in a more uniform gradient of $\mathrm{H\delta_A}+\mathrm{H\gamma_A}$ across the plane, especially at the bright end.

These results indicate that the aperture corrections reduce the noise in the correlations between structural and stellar populations parameters.
\end{appendix}
\end{document}